\documentclass[aps,prd,twocolumn,showpacs,floats,floatfix,letterpaper,nofootinbib,superscriptaddress,]{revtex4-1}

\usepackage{amssymb,amsmath,latexsym,mathrsfs}
\usepackage{graphicx,subfigure}
\usepackage{epsfig}
\usepackage{varioref,xr-hyper}
\usepackage{color}
\usepackage{multirow}
\usepackage{array}

\newcommand{\neff}{N_{\rm {eff}}}

\newcommand{\nnusa}{N_{\rm {\nu}}^{4}}

\newcommand{\summnu}{\sum m_{\nu}}

\newcommand{\msa}{m_{4}}
\newcommand{\msb}{m_{5}}

\begin{document}

\title{Sterile Neutrinos: Cosmology vs Short-BaseLine Experiments}

\author{Maria Archidiacono}
\affiliation{Department of Physics and Astronomy, Aarhus University, 8000 Aarhus C, Denmark}

\author{Nicolao Fornengo}
\affiliation{Department of Physics, University of Torino and INFN, Via P. Giuria 1, I--10125 Torino, Italy}

\author{Carlo Giunti}
\affiliation{Department of Physics, University of Torino and INFN, Via P. Giuria 1, I--10125 Torino, Italy}

\author{Steen Hannestad}
\affiliation{Department of Physics and Astronomy, Aarhus University, 8000 Aarhus C, Denmark}

\author{Alessandro Melchiorri}
\affiliation{Physics Department, Universit\`a di Roma ``La Sapienza'' and INFN, P.le Aldo Moro 2, 00185, Rome, Italy}

\begin{abstract}
Cosmology and short baseline neutrino oscillation data both hint at the existence of light sterile neutrinos with masses in the 1 eV range. Here we perform a detailed analysis of the sterile neutrino scenario using both cosmological and SBL data. We have additionally considered the possibility that the extra neutrino degrees of freedom are not fully thermalised in the early universe. Even when analyzing only cosmological data we find a preference for the existence of massive sterile neutrinos in both (3+1) and (3+2) scenarios, and with the inclusion of SBL data the evidence is formally at the 
3.3$\sigma$ level in the case of a (3+1) model.
 Interestingly, cosmological and SBL data both point to the same mass scale of approximately 1 eV. 
In the (3+1) framework WMAP9+SPT provide a value of the sterile mass eigenstate $\msa = (1.72\pm0.65)$ eV:
this result is strenghtened by adding the prior from SBL posterior to $\msa = (1.27\pm0.12)$ eV
($\msa = (1.23\pm0.13)$ eV when SDSS is also considered in the cosmological analysis).
In the (3+2) scheme, two additional, non--fully thermalized, neutrinos are compatible with
the whole set of cosmological and SBL data, leading to mass values of $\msa = (0.95\pm0.30)$ eV
and $\msb = (1.59\pm0.49)$ eV.
The inclusion of Planck data does not change our considerations about the mass scale;
concerning the extra neutrino degrees of freedom, invoking a partial thermalisation the 3+1 model is still consistent with the latest data.
\end{abstract}


\maketitle

\section{Introduction}

The past decade has seen a rapid increase in our understanding of neutrinos. Oscillation experiments have now established the main structure of the leptonic mixing matrix and provided evidence for at least two neutrino mass states of non-zero mass (see e.g. Ref. \cite{Fogli:2012ua} for a recent overview). However, some crucial questions are still left unanswered: the absolute mass scale of neutrinos is extremely hard to measure in laboratory experiments and is therefore currently only poorly known; the hierarchy of neutrino masses is not yet disentangled; the possible existence of additional neutrino species beyond the three predicted by the Standard Model is still an open issue.
In fact, short--baseline (SBL) oscillation experiments, as well as reactor neutrino flux measurements seem to hint at the existence of a
fourth (3+1 models, possibly with non standard interactions) or fifth (3+2 models) sterile neutrinos, mixing with the active neutrinos and with a mass around 1 eV,
see Refs. \cite{Kopp:2011qd,Giunti:2011gz,Giunti:2011hn,Giunti:2011cp,Karagiorgi:2012kw,Donini:2012tt}
.

Cosmology has, at the same time, provided important insights into some of these questions. Neutrinos are produced in copious amounts in the early Universe, and are still present as a cosmic neutrino background. Even though this background is extremely difficult if not impossible to measure directly, it influences a number of cosmological observables such as the cosmic microwave background (CMB), and the power spectrum of matter fluctuations.
The effective number of relativistic degrees of freedom changes both the position and the shape of peaks of the CMB temperature power spectrum at high multipoles \cite{Hou:2011ec}. The neutrino mass affects both the CMB, through the enhancement of the early integrated Sachs--Wolfe effect, and the matter power spectrum, via the free--streaming that suppresses the power on small scales. Thanks to these fingerprints, cosmology can strongly constrain the absolute neutrino mass scale and the cosmic neutrino background \cite{Hannestad:2007tu, Hamann:2012fe}.

In the past years cosmology has provided some hints for a non standard value of the effective number of relativistic degrees of freedom,
see Refs. \cite{Giusarma:2011ex, Giusarma:2011zq, Hamann:2011zz, Hamann:2012fn}, with the preferred value in the late--time universe (around, or subsequent to, recombination) being significantly higher than the standard model prediction of $\neff = 3.046$ \cite{Mangano:2005cc}. Such additional relativistic energy density is usually referred to as dark radiation and can arise from a completely different physics such as axions \cite{Marsh:2011bf}.
The light sterile neutrinos hinted at by SBL data would be an excellent candidate for dark radiation \cite{Archidiacono:2011gq}, even though misinterpretations of the nature of this non standard $\neff$ can arise from degeneracies between $\neff$ and other cosmological parameters \cite{Archidiacono:2012gv}.

Recently, new CMB data released by the WMAP collaboration \cite{wmap9}, the South Pole Telescope (SPT) \cite{spt}, and the Atacama Cosmology Telescope (ACT) \cite{act} have led to a somewhat confusing situation concerning dark radiation
and, in general, those neutrino properties that can be constrained by cosmology:
the WMAP9 and SPT data both confirm the presence of an extra dark radiation component, while the new ACT data seem pointing towards a value of $\neff$ in agreement with the standard model prediction,
in contrast with their own previous analysis \cite{Dunkley:2010ge}.

Fortunately this problem might be resolved by new data from the Planck mission \cite{PLANCK} which should be able to constrain $\neff$ with a much better precision than existing data (see e.g. Ref. \cite{Perotto:2006rj,Hamann:2007sb}).
However, even if the incoming cosmological data from Planck confirms the standard cosmological value for $\neff$, cosmology should address the discrepancy with the SBL neutrino oscillation results. Since the SBL experiments do not provide direct constraint on the cosmological thermalization of the sterile neutrinos, a small lepton asymmetry can be invoked to reduce the thermalization efficiency (see e.g. Ref. \cite{Hannestad:2012ky,Saviano:2013ktj} for recent treatments) and reconcile these two branches of neutrino physics.

So it is indeed timely to investigate the interplay between neutrino oscillation experiments and cosmology in determining the neutrino properties \cite{Archidiacono:2012ri, Joudaki:2012uk, Jacques:2013xr, Kristiansen:2011mp, Melchiorri:2008gq}.
This paper is aimed at investigating the joint constraints on neutrino number and neutrino mass from these two different branches.
The paper is organized as follows: in Section \ref{sec:SBL} we present the SBL analysis data, which
includes updates on our previous analysis; in Section \ref{sec:cosmologicalmethod} we describe the cosmological data used in this work and the method we have applied to analyze them; in Section \ref{sec:cosmologicalresults} we provide an update of the cosmological constraints on  sterile neutrinos; in Section \ref{sec:jointanalysis} we perform the joint analysis by applying a short--baseline prior to the cosmological analysis.
Section \ref{sec:conclusions} contains our conclusions.
Finally in Section \ref{sec:planck} we have applied the same model of \ref{subsec:3p1jointanalysis} to the analysis of Planck data \cite{Ade:2013lta} in order to check that the latest CMB measurements do not change our conclusions.

\section{SBL Analysis}
\label{sec:SBL}

We consider 3+1 and 3+2 extensions of the standard three--neutrino mixing
(see Ref. \cite{GonzalezGarcia:2007ib})
in which we have one ($m_{4}$)
or two  ($m_{4}$ and $m_{5}$)
new neutrino masses at the eV scale
and the masses of the three standard neutrinos are much smaller:
\begin{equation}
m_{1}
,\,
m_{2}
,\,
m_{3}
\ll
m_{4}
\leq
m_{5}
\,.
\label{hierarchy}
\end{equation}
In this case,
the squared-mass differences
\begin{equation}
\Delta{m}^2_{41} \simeq m_{4}^2
\quad
\text{and}
\quad
\Delta{m}^2_{51} \simeq m_{5}^2
\,,
\label{dm2}
\end{equation}
with
$\Delta{m}^2_{kj} \equiv m_{k}^2 - m_{j}^2$,
generate short--baseline oscillations
through the mixing relation
\begin{equation}
\nu_{\alpha}
=
\sum_{k=1}^{4\,\text{or}\,5} U_{\alpha k} \nu_{k}
\,,
\label{mixing}
\end{equation}
between the flavor neutrino fields $\nu_{\alpha}$
($\alpha=e,\mu,\tau,s_{1},s_{2}$ in the 3+2 model)
and
the neutrino fields $\nu_{k}$ with masses $m_k$
($k=1,\ldots,5$ in the 3+2 model).
$U$ is the unitary mixing matrix.

In the analysis of short--baseline data we adopted the approach described in Ref.
\cite{Archidiacono:2012ri},
with some improvements in the considered data sets,
which can be divided in the following three groups:

\begin{enumerate}

\item
The
$\nu_{\mu}\to\nu_{e}$ and $\bar\nu_{\mu}\to\bar\nu_{e}$
appearance data of the
LSND \cite{Aguilar:2001ty},
KARMEN \cite{Armbruster:2002mp},
NOMAD \cite{Astier:2003gs},
MiniBooNE \cite{AguilarArevalo:2012va} and
ICARUS \cite{Antonello:2012pq}
experiments.
In particular,
we use only the MiniBooNE data above 475 MeV,
because the data at lower energy contains an anomaly which
cannot be explained by neutrino oscillations
\cite{Giunti:2011hn,Giunti:2011cp}
(an interesting possiblity of reconciling the low--energy anomalous data with neutrino oscillations
through energy reconstruction effects has been suggested and discussed in Refs.
\cite{Martini:2012fa,Martini:2012uc}).

\item
The
$\nu_{e}$ and $\bar\nu_{e}$
disappearance data described in Ref. \cite{Giunti:2012tn},
which take into account the
reactor
\cite{Mention:2011rk}
and
Gallium
\cite{Giunti:2010zu}
anomalies.

\item
The constraints on
$\nu_{\mu}$ and $\bar\nu_{\mu}$
disappearance obtained from
the data of the
CDHSW experiment \cite{Dydak:1983zq},
from the analysis \cite{Maltoni:2007zf} of
the data of
atmospheric neutrino oscillation experiments,
from the analysis \cite{Giunti:2011hn} of the
MINOS neutral-current data \cite{Adamson:2011ku}
and from a new analysis of the
SciBooNE-MiniBooNE
neutrino \cite{Mahn:2011ea} and antineutrino \cite{Cheng:2012yy} data.

\end{enumerate}

\begin{table}[t]
\begin{center}
\begin{tabular}{|c|c|c|c|}
\hline
&
{\bf 3+0}
&
{\bf 3+1}
&
{\bf 3+2}
\\
\hline
$\chi^2_{\text{min}}$			& $280.2	$ & $236.1	$ & $229.0	$ \\
$\text{NDF}$				& $230		$ & $227	$ & $223	$ \\
$\text{GoF}$				& $1.3\%	$ & $32\%	$ & $38\%	$ \\
\hline
$\Delta{m}^2_{41}\,[\text{eV}^2]$	& $		$ & $1.62	$ & $0.91	$ \\
$|U_{e4}|^2$				& $		$ & $0.031	$ & $0.015	$ \\
$|U_{\mu4}|^2$				& $		$ & $0.01	$ & $0.011	$ \\
$\Delta{m}^2_{51}\,[\text{eV}^2]$	& $		$ & $		$ & $1.61	$ \\
$|U_{e5}|^2$				& $		$ & $		$ & $0.0226	$ \\
$|U_{\mu5}|^2$				& $		$ & $		$ & $0.00664	$ \\
$\eta$					& $		$ & $		$ & $1.56\pi	$ \\
\hline
$\Delta\chi^2_{\text{PG}}$		& $		$ & $6.6	$ & $11.12	$ \\
$\text{NDF}_{\text{PG}}$		& $		$ & $2		$ & $4		$ \\
$\text{GoF}_{\text{PG}}$		& $		$ & $4\%	$ & $2.5\%	$ \\
\hline
\end{tabular}
\end{center}
\caption{ \label{tab:sbl}
Values of
$\chi^{2}_{\text{min}}$,
number of degrees of freedom (NDF),
goodness--of--fit (GoF)
and
best--fit values of the mixing parameters
obtained in our 3+0, 3+1 and 3+2 fits of short--baseline oscillation data.
The last three lines give the appearance--disappearance
parameter goodness--of--fit (PG).
}
\end{table}

The results of a least--squares analysis of the SBL data is
presented in Tab.~\ref{tab:sbl}.
First, we notice that the values of the goodness--of--fit
obtained in the 3+1 and 3+2 models
are satisfactory and much better than what is
obtained in the case of absence of sterile neutrinos (3+0).
We notice also that the value of the appearance--disappearance
parameter goodness--of--fit is acceptable
and it is remarkable that it is better
in the 3+1 model than in the 3+2 model,
contrary to previous results with older data
\cite{Kopp:2011qd,Giunti:2011gz,Conrad:2012qt}.
The reason is that
the values of $\chi^2_{\text{min}}$
for appearance (APP) and disappearance (DIS) data
in the two models are:
\begin{align}
\null & \null
(\chi^2_{\text{min}})^{\text{3+1}}_{\text{APP}} = 50.4
,\,
\null && \null
(\chi^2_{\text{min}})^{\text{3+1}}_{\text{DIS}} = 179.1
\,,
\label{chi2min3p1}
\\
\null & \null
(\chi^2_{\text{min}})^{\text{3+2}}_{\text{APP}} = 40.1
,\,
\null && \null
(\chi^2_{\text{min}})^{\text{3+2}}_{\text{DIS}} = 177.8
\,.
\label{chi2min3p2}
\end{align}
Since the 3+2 model can fit significantly better than the 3+1 model only the appearance data,
the larger difference between the global $\chi^2_{\text{min}}$
and
$(\chi^2_{\text{min}})_{\text{APP}} + (\chi^2_{\text{min}})_{\text{DIS}}$
in the 3+2 model cannot be compensated by the increase of the number of degrees of freedom.

\begin{figure}[t]
\begin{center}
\includegraphics[bb=5 21 560 567, width=\linewidth]{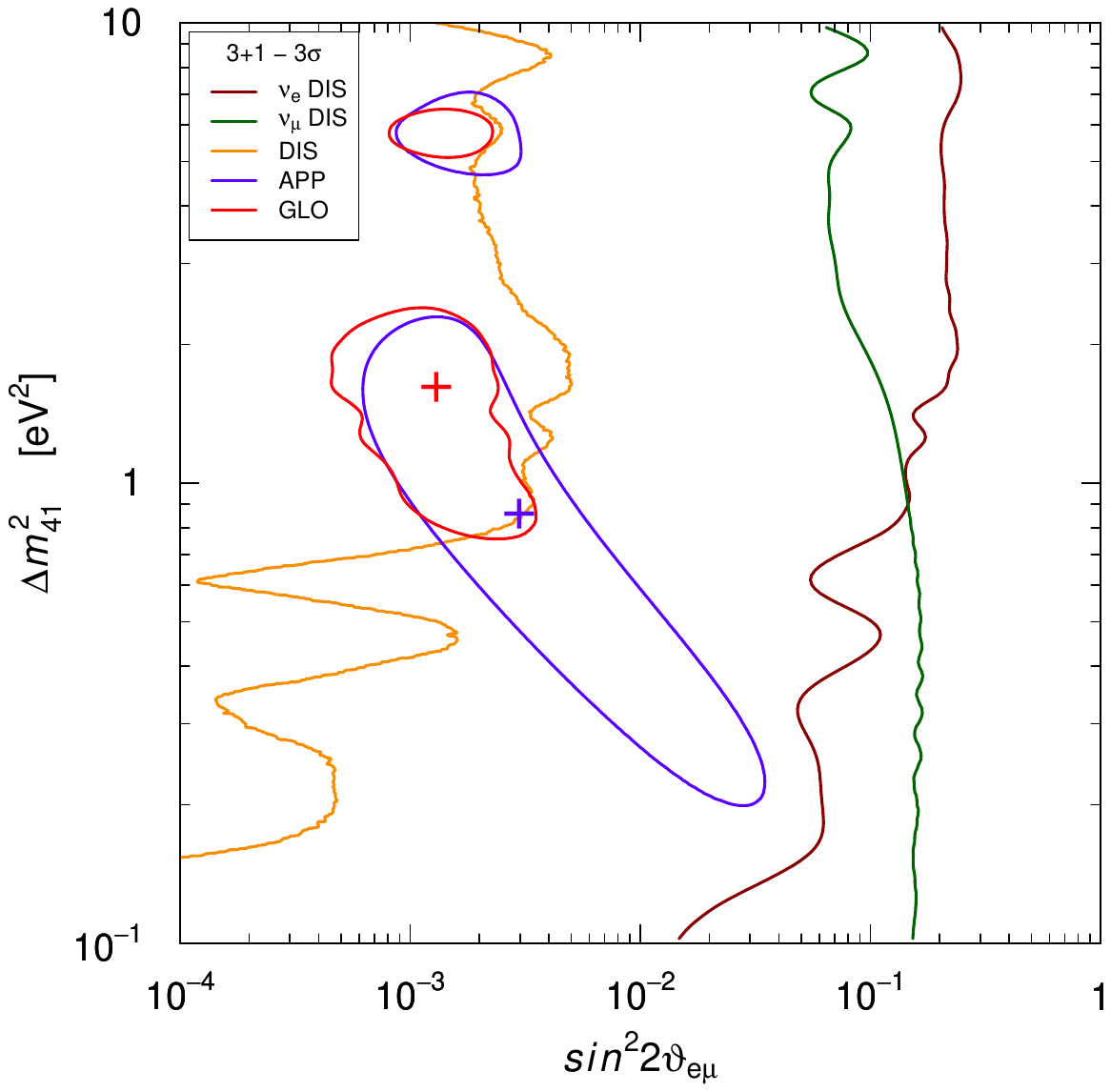}
\end{center}
\caption{ \label{fig:sbl}
Allowed $3\sigma$ regions (99.73\% CL) in the
$\sin^{2}2\vartheta_{e\mu}$--$\Delta{m}^{2}_{41}$ plane
in the 3+1 model obtained from
$\nu_{e}$ and $\bar\nu_{e}$
disappearance data (left of the dark red curve),
$\nu_{\mu}$ and $\bar\nu_{\mu}$
disappearance data (left of the dark green curve),
combined disappearance data (left of the dark orange curve),
$\nu_{\mu}\to\nu_{e}$ and $\bar\nu_{\mu}\to\bar\nu_{e}$
appearance data (inside the blue curve)
and from the global fit (inside the red curves).
The best-fit points in the last two cases are indicated by crosses.
}
\end{figure}

Figure~\ref{fig:sbl}
shows the allowed $3\sigma$ regions (99.73\% CL) in the
$\sin^{2}2\vartheta_{e\mu}$--$\Delta{m}^{2}_{41}$ plane
in the 3+1 model
obtained with disappearance data,
appearance data, and all data respectively.
The quantity $\sin^{2}2\vartheta_{e\mu} = 4 |U_{e4}|^2 |U_{\mu4}|^2$
is the amplitude of
$\nu_{\mu}\to\nu_{e}$ and $\bar\nu_{\mu}\to\bar\nu_{e}$
oscillations
(see Ref. \cite{GonzalezGarcia:2007ib}).
One can see that
the combination of the constraints obtained from
disappearance data excudes the large-$\sin^{2}2\vartheta_{e\mu}$
part of
the regions allowed by appearance data.
From the global fit we obtained
a low--$\Delta{m}^{2}_{41}$
allowed region at
$\Delta{m}^{2}_{41} \simeq 0.8 - 2 \, \text{eV}^2$
and
$\sin^{2}2\vartheta_{e\mu} \simeq (0.5 - 3) \times 10^{-3}$,
containing the best--fit point,
and a high--$\Delta{m}^{2}_{41}$
allowed region at
$\Delta{m}^{2}_{41} \simeq 6 \, \text{eV}^2$
and
$\sin^{2}2\vartheta_{e\mu} \simeq (0.8 - 2) \times 10^{-3}$.

The appearance--disappearance tension discussed in previous papers
(e.g. Refs. \cite{Kopp:2011qd,Giunti:2011gz,Giunti:2011hn,Giunti:2011cp,Conrad:2012qt})
is still present, because,
as one can see from Fig.~\ref{fig:sbl},
the best--fit point of the appearance data
is excluded at about $3\sigma$ by the disappearance data.
However,
the tension is less severe than that which was obtained with old data,
as testified by the acceptable parameter goodness--of--fit
in Tab.~\ref{tab:sbl}.

In the following we combine the results of
the analysis of short--baseline and cosmological data
considering both 3+1 and 3+2 schemes,
in spite of the fact that the 3+1 scheme is sufficient
to explain the current short--baseline data and
the 3+2 scheme is disfavored by Occam's razor.
We consider also the 3+2 scheme because of the current interest in it
(see, for example, Ref. \cite{Abazajian:2012ys})
and because future data may reverse the preference.

\section{Cosmological Method}
\label{sec:cosmologicalmethod}

We have modified the Monte Carlo Markov Chain (MCMC) public package \texttt{CosmoMC} \cite{Lewis:2002ah} (October 2012 version) to account for the datasets listed below and in order to sample the extended parameter space for our (3+1) and (3+2) investigations.

The cosmological analysis is performed by employing various combinations of datasets.
The WMAP 9--years data release \cite{wmap9} represents our basic CMB dataset. At high multipoles, we use new data from the CMB experiments South Pole Telescope \cite{spt} and Atacama Cosmology Telescope \cite{act}.
The information on dark matter clustering comes from the matter power spectrum derived from
the Sloan Digital Sky Survey Data Release 7 luminous red galaxy sample \cite{red}.
We have also investigated the impact on our constraints of the
Baryonic Acoustic Oscillations (BAO) results of Refs. \cite{Anderson:2012sa, Padmanabhan:2012hf, Blake:2011en} and of the prior on the Hubble constant coming from the Hubble Space Telescope measurements \cite{hst}.

Short--baseline results have been included as a prior in the analysis.
Therefore the final $\chi^2$ of the combined analysis is simply given by
\begin{equation}
\chi^2_{\rm{tot}}=\chi^2_{\rm{cosmology}}+\chi^2_{\rm{SBL}}.
\end{equation}
Concerning the inclusion of the SBL information in the cosmological analyses,
the difference with respect to our previous work \cite{Archidiacono:2012ri},
is that here we directly incorporate the $\chi^2$ coming from SBL data (analyzed in a variety of frameworks)
in the MCMC sampling: this allows for unbiased constraints on those parameters that
are common to the cosmological and SBL analyses.

Our basic cosmological model is the 6--parameters flat $\Lambda$CDM model. Its
parameters are:
the physical baryon and cold dark matter densities
$\Omega_{b} h^2$ and $\Omega_{dm} h^2$, the ratio of the sound horizon to the angular
diameter distance at decoupling $\theta_s$, the optical depth to reionization $\tau$,
the scalar spectral index $n_s$ and the overall normalization of the spectrum $A_s$.
In the cosmological analyses, we also account for the Sunyaev--Zeldovich (SZ) effect and foreground contributions by including three extra amplitudes: the SZ amplitude $a_{SZ}$,
the amplitude of clustered point sources $a_{c}$
and the amplitude of Poisson distributed point--sources $a_{p}$. Since the ACT team uses a different mask for identifying and removing point sources, when the ACT data are  included in our analysis (Section \ref{sec:cosmologicalresults}) a fourth extra amplitude is needed to account for the Poisson contribution in the ACT data $a_{pACT}$ different from $a_{pSPT}$. Furthermore when the ACT likelihood is included in our analysis, we split the SZ contribution in two amplitudes referred for the two different SZ effects: $a_{kSZ}$ for the kinetic SZ effect and $a_{tSZ}$ for the thermal SZ effect.

In order to include extra sterile neutrinos, the $\Lambda$CDM model is extended to a $\Lambda$ Mixed Dark Matter model ($\Lambda$MDM) by introducing a hot dark matter component in the form of massive neutrinos. Usually this component is parameterized as the neutrino mass fraction $f_\nu$ (the neutrino--hot dark matter density over the total dark matter density),
as in Section \ref{sec:cosmologicalresults},
but in the joint analysis of Section \ref{sec:jointanalysis} the masses of the single mass eigenstate are used as free parameters, in order to directly sample the parameter which the SBL prior acts on.

With the above assumptions, the contribution of massive neutrinos to the energy budget of the Universe follows the
usual relation:
\begin{equation}
\Omega _{\nu} h^2=\frac{\sum m_\nu}{93.14 \: \rm{ eV}}.
\label{eq:omega}
\end{equation}
where $\sum m_\nu$ denotes an effective sum of masses, specified below for the different
cases under study.

Before attempting the combined (cosmological + SBL) analysis, we
perform a number of tests on the cosmological data alone, in order to assess the requirements
on dark radiation from the various cosmological data sets. These results will then be compared
to the interpretation of dark radiation in terms of sterile neutrinos in the (3+1) and (3+2) models.
From the point of view of cosmology there is no difference between active neutrinos and sterile neutrinos provided they are fully thermalized. For the typical mass differences and mixing angles hinted at by the SBL data this is indeed the case (see e.g.\ \cite{Hannestad:2012ky,Saviano:2013ktj}). In this case only the total sum of neutrino masses and the total (effective) number of neutrino species are relevant parameters (with the current level of precision of cosmological data). However, it is entirely possible that the sterile states are only partially thermalized. This can for example happen in models with non-zero lepton asymmetry \cite{Hannestad:2012ky,Saviano:2013ktj}. Therefore, in Section \ref{sec:cosmologicalresults} we perform three different analyses:

\begin{itemize}
\item[(A)] All neutrinos are massless: in this case cosmological observables are just sensitive
to the total effective number of relativistic (fermionic) degrees of freedom denoted by $N_{\rm eff}$,
where $\neff = 3.046$ \cite{Mangano:2005cc} if only (effectively massless) standard model neutrinos
are present;

\item[(B)] Neutrinos are allowed to be massive, with a common total mass $\summnu$ and
a number effectively equal to $N'_{\rm eff}$. This corresponds to:
\begin{equation}
\sum m_\nu = N'_{\rm eff} \times m_M
\end{equation}

\item[(C)] The three standard model neutrinos are (effectively)
massless, while extra sterile neutrinos are present in (effective) number equal to $N_S$ and total (effective) mass equal to $m_S$. This corresponds to:
\begin{equation}
\sum m_{\nu S} = N_S \times m_S
\end{equation}
and to a total effective number of neutrinos $N_{\rm eff} = 3.046 + N_S$.
This case differs from case (B) since here the presence of 3 massless neutrinos
fully contributes to radiation energy, while in case (B) all neutrinos are effectively
massive and there may be a different impact on matter-radiation equality, CMB anisotropies
and cosmic structures formation.
 Clearly in this case $N_S$ measures only the extra-number of degree of freedom responsible
for dark radiation, while in case (B) $N'_{\rm eff}$ is the total number of (potentially) massive
neutrinos.

\end{itemize}

We have considered the possibility that sterile neutrinos are only partially thermalised. In our analysis we express this in terms of the following parameter:
\begin{equation}
N_i = \frac{n_i}{n_i^{\rm th}}
\end{equation}
where $n_i$ denotes the actual number density of a sterile neutrino
of mass $m_i$ while $n_i^{\rm th}$ denotes the number density of a standard neutrino (fully thermalized, with a Fermi--Dirac phase--space distribution) with the same mass $m_i$. The
multiplicity parameter $N_i$ therefore defines the fractional contribution of the (non--standardly thermalized)
sterile neutrino to the dark matter energy density, and is a number defined in the interval $[0,1]$.
In our assumptions, $N_i$ is considered to be independent from the mass $m_i$, and therefore
$N_i$ and $m_i$ can be treated as independent parameters in the analysis. While this may not
be the most general case, it is generic enough to study the possibility to allow, in the cosmological
data, for extra sterile neutrinos endowed
with the properties dictated by the SBL studies. Examples of values of $N_i < 1$ may be related
to partial thermalization (possibly related to the presence of a lepton asymmetry, since the mixing
angles obtained from the SBL analysis are large enough to ensure thermalization of the sterile neutrinos
if the asymmetry is absent) or non--standard phase--space distributions for sterile neutrinos. The limit
$N_i=1$ refers to a fully thermalized Fermi--Dirac sterile neutrino.

In out joint (cosmological + SBL) analysis (discussed in Section \ref{sec:jointanalysis}) we
assume the three active neutrinos as massless and the $\sum m_\nu$ in Eq. (\ref{eq:omega}) is therefore given by $\sum m_\nu = \sum_i N_i m_i$, where the index $i$
runs over the number of sterile neutrinos considered in our analysis:
\begin{align}
\null & \sum m_\nu = N_4 \times m_4   & \mbox{(3+1)} \label{eq:(3+1)}\\
\null & \sum m_\nu = N_4 \times m_4 + N_5 \times m_5  & \mbox{(3+2)} \label{eq:(3+2)}
\end{align}
We have considered a top--hat prior in the range $[0,1]$ for the multiplicity parameters, while the sterile neutrino masses have been subjected to a prior given by the $\chi^2$ results of the SBL analysis.

Finally, in order to grasp a connection between the joint (cosmological + SBL) analysis and the more
typical cosmological investigations (where the relevant parameters are just the sum of the neutrino masses and an effective number of additional light states, as discussed above), we have re--analyzed the  (3+2) case by using the SBL--prior on $m_4$ and $m_5$ projected over the sum of the two masses
$m_4 + m_5$. This case corresponds again to a situation where:
\begin{align}
\null & \sum m_\nu = N_4 \times m_4 + N_5 \times m_5  & \mbox{(3+2)}
\end{align}
where we look at the results in terms of $(N_4+N_5)$ and $m_4 + m_5$.
This case is discussed in Section \ref{sec:jointanalysis} and differs from the case studied in connection
with case (C) because information coming from the SBL studies is included in the analysis.

\section{Cosmological Results}
\label{sec:cosmologicalresults}

\begin{table*}[t]
\begin{center}
\begin{tabular}{|l|c|c|c|}
\hline
\hline
 Parameters & Case (A) & Case (B) & Case (C)\\
                   & WMAP9+SPT+ACT                & WMAP9+SPT+ACT                      &WMAP9+SPT+ACT \\
\hline
$\Omega_b h^2$ & $0.02259\pm0.00045$ & $0.02224\pm0.00053$ & $0.02281\pm0.00035$ \\
$\Omega_{dm} h^2$ & $0.115\pm0.008$ & $0.120\pm0.009$ & $0.126\pm0.009$ \\
$\theta_s$ & $1.0418\pm0.0013$ & $1.0417\pm0.0013$ & $1.0416\pm0.0011$ \\
$\tau$ & $0.086\pm0.014$ & $0.086\pm0.013$ & $0.088\pm0.013$ \\
$n_s$ & $0.972\pm0.017$ & $0.959\pm0.020$ & $0.966\pm0.010$ \\
$\log(10^{10} A_s)$ & $3.178\pm0.041$ & $3.213\pm0.049$ & $3.212\pm0.039$ \\
\hline
$\neff$ \hfill (A)& $3.17\pm0.47$ & $-$ & $-$ \\
\hline
$N'_{\rm eff}$\hfill (B) & $-$ & $2.97\pm0.48$ & $-$ \\
$\summnu [{\rm eV}]$\hfill (B) & $-$ & $<1.17$ & $-$ \\
\hline
$N_S$ \hfill(C) & $-$ & $-$ & $<0.89$ \\
$\Sigma m_{\nu S} [{\rm eV}]$ \hfill(C) & $-$ & $-$ & $<2.39$ \\
\hline
$H_0 [\rm{km/s/Mpc}]$ & $71.7\pm3.4$ & $64.9\pm5.5$ & $68.9\pm2.0$ \\
$\sigma_8$ & $0.822 \pm 0.029$ & $0.714\pm0.073$ & $0.699\pm0.071$ \\
$\Omega_{\rm m}$ & $0.269 \pm 0.020$ & $0.345\pm0.063$ & $0.315\pm0.034$ \\
$a_{tSZ}$ & $4.4\pm0.9$ & $4.4\pm0.9$ & $4.5\pm0.9$ \\
$a_{kSZ}$ & $<3.2$ & $<3.4$ & $<3.7$ \\
$a_c$ & $6.0\pm0.5$ & $6.0\pm0.5$ & $6.2\pm0.5$ \\
$a_{pSPT}$ & $18.5\pm1.6$ & $18.6\pm1.6$ & $18.4\pm1.6$ \\
$a_{pACT}$ & $7.0\pm0.3$ & $6.9\pm0.3$ & $6.9\pm0.3$ \\
\hline
$\chi^2_{\rm min}$ & $8962.2$ & $8962.9$ & $8961.3$ \\
\hline
\hline
\end{tabular}
\vspace{0.3cm}
\caption{Values of the cosmological parameters and their $68 \%$ confidence level intervals
for the the three cosmological analysis described in Section \ref{sec:cosmologicalmethod}:
case (A) refers to $N_{\rm eff}$ massless neutrinos; case (B) refers to $N'_{\rm eff}$ massive
neutrinos with total mass $\summnu$; case (C) refers to 3 massless (active) neutrinos
plus $N_S$ additional (sterile) neutrinos, with total mass $m_S$. Upper bounds are quoted at $95 \%$ C.L.}
\label{tab:cosmo}
\end{center}
\end{table*}

\begin{figure}[t]
\includegraphics[scale=0.45]{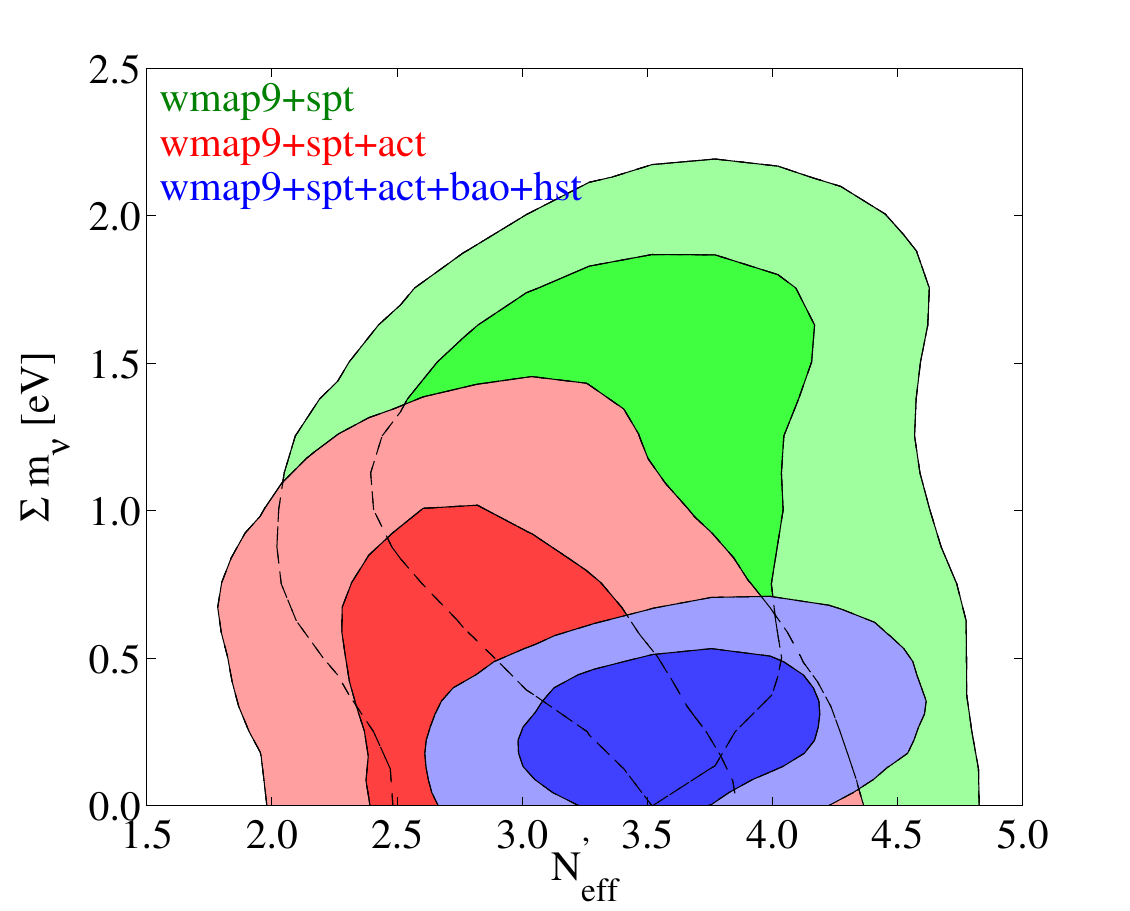}
\caption{Case (B): $N'_{\rm eff}$ massive
neutrinos with total mass $\summnu$ -- Two dimensional marginalized 68\% and 95\% confidence level regions in the plane $\summnu$ vs $N'_{\rm eff}$ .
Leftmost (red) contours refer to CMB--only data (WMAP9+SPT+ACT), while rightmost (blue) contours include also BAO and HST. The larger (green) area denotes the results for WMAP9+SPT
datasets.}
\label{fig:cosmo}
\end{figure}

\begin{figure*}[t]
\includegraphics[scale=0.45]{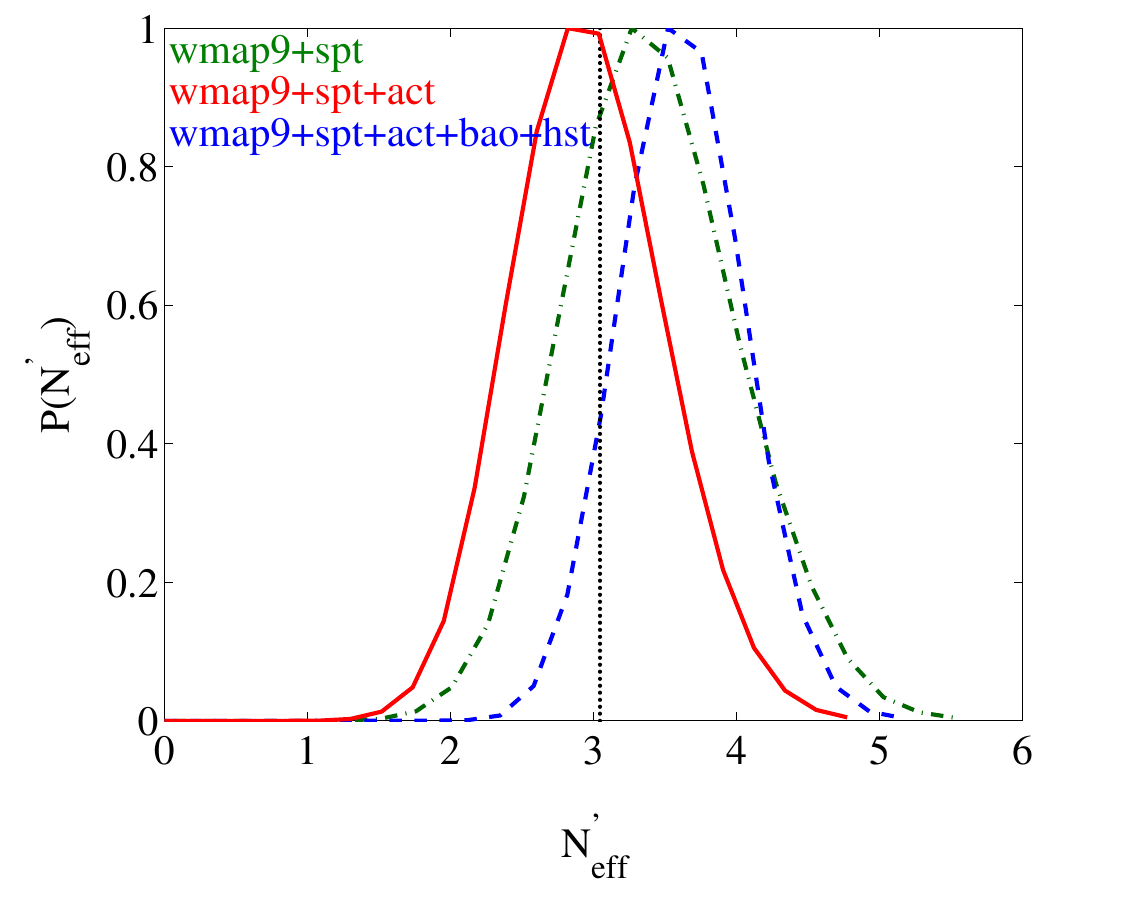}
\includegraphics[scale=0.45]{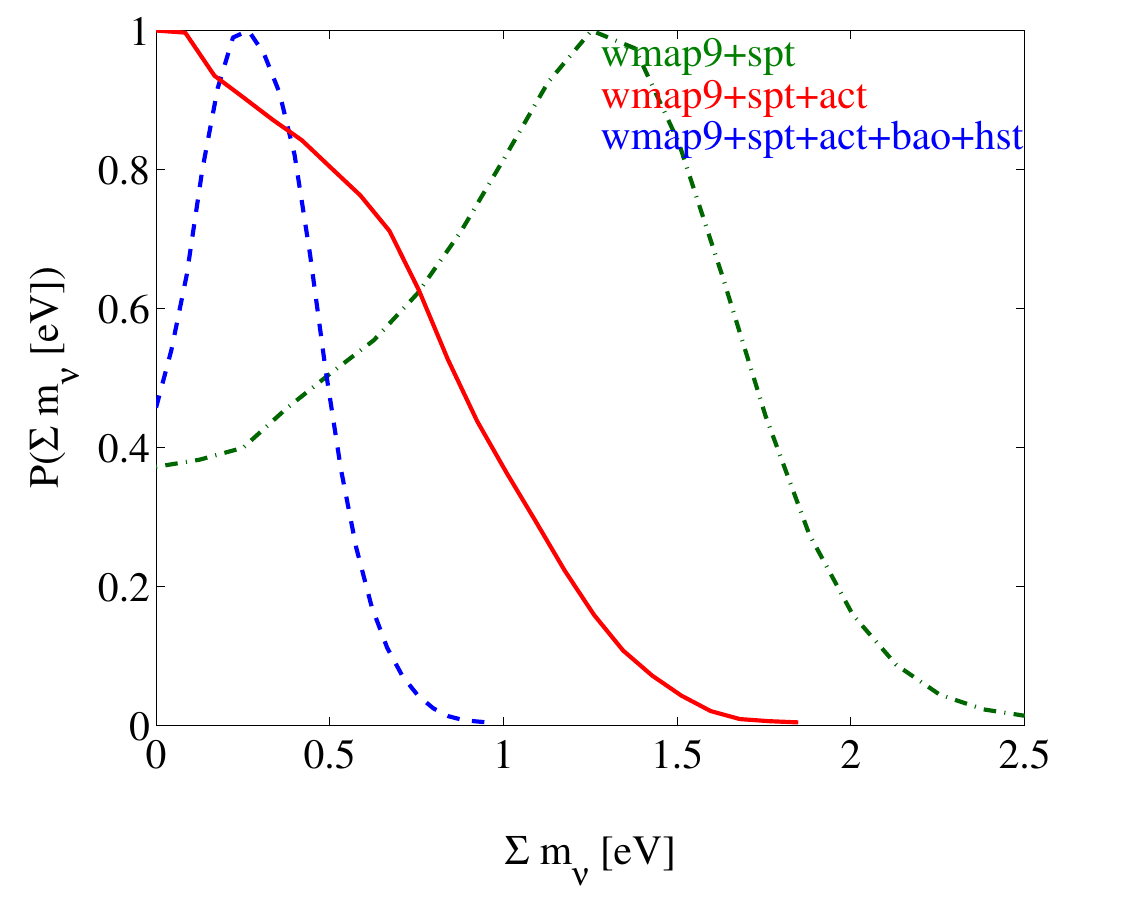}
\caption{Case (B): $N'_{\rm eff}$ massive
neutrinos with total mass $\summnu$ -- One dimensional marginalized posteriors for $N'_{\rm eff}$ (left panel) and $\summnu$ (right panel).
Solid (red) lines refer to CMB--only data (WMAP9+SPT+ACT), while dashed (blue) lines include also BAO and HST. The dot--dashed (green) curve denotes the results for WMAP9+SPT
datasets.}
\label{fig:cosmo2}
\end{figure*}

The cosmological analyses performed on the full CMB dataset (WMAP9, SPT and ACT combined) is shown in Table \ref{tab:cosmo}. The results for case (B) are also illustrated in Figure \ref{fig:cosmo}, where the marginalized 68\% and 95\% contours for $N'_{\rm eff}$ and $\summnu$ are reported. The analysis
for the combination of  WMAP9+SPT+ACT datasets refers to the leftmost (red) contours; the
further inclusion of BAO+HST moves the contours to the right (blue contours), referring to larger $N'_{\rm eff}$, accompanied by smaller values of $\summnu$). Figures \ref{fig:cosmo2} shows the one dimensional marginalized posteriors for $N'_{\rm eff}$ and $\summnu$.

The results shown in Table \ref{tab:cosmo} and Figures \ref{fig:cosmo} and \ref{fig:cosmo2} point
toward the conclusion that the full set of present CMB data do not show any indication for a non--standard value of the effective number of relativistic degrees of freedom.
Nevertheless this result is mostly driven by the new ACT likelihood and a strong tension actually emerges between the new SPT and the new ACT data. This tension turns out in a bias that has a strong impact on the cosmological results, especially on the neutrino sector. Removing the ACT data and using only WMAP9 and SPT, in the massless neutrino case (case (A)) we recover a $1.5 \sigma$ evidence for an extra dark radiation component:
\begin{align*}
&N_{\rm eff} = 3.87\pm0.55 \: (1\sigma) && \mbox{case (A)} \\
&\mbox{WMAP+SPT} && ~ \nonumber
\end{align*}
This result is fully consistent with those of the WMAP9 \cite{wmap9} and SPT \cite{spt} analyses.
In case of massive neutrinos (case (B)) Figure \ref{fig:cosmo} shows that, when all CMB data are considered (red contours), an anti--correlation emerges between $N'_{\rm eff}$ and the total mass $\summnu$: a higher value of the sum of the masses seems to be consistent with a lower value of the number of neutrinos that share the same mass. This inconsistency can be traced to the tension between the high multipole datasets (ACT and SPT). Indeed, excluding the ACT data, we get the green conturs of Figure \ref{fig:cosmo}: the higher mass values allowed by SPT correspond to higher numbers of neutrino species, and there isn't any strong correlation between these two quantities. Furthermore the above tension between ACT and SPT is less pronounced also if we take into account BAO and HST data. In this case (blue contours) we recover the expected positive correlation between $\summnu$ and $N'_{\rm eff}$,
because the HST measurements fix the expansion rate and, as a consequence, the dimension of the sound horizon at recombination and the damping scale. So an enhancement in the sum of the masses is needed in order to get a higher value of the number of neutrino species.
Moreover, considering WMAP9+SPT+ACT+BAO+HST, we get a slight preference for a non--standard value of the effective number of relativistic degrees of freedom (see Figure \ref{fig:cosmo2}):
\begin{align*}
& N'_{\rm eff}  =  3.60 \pm 0.35 \: (1\sigma) && \mbox{case (B)} \\
& \mbox{WMAP+SPT+ACT+BAO+HST} && ~ \nonumber
\end{align*}
Concerning the mass, the constraint on $\summnu$ is significantly tightened by adding the low redshift observations (BAO and HST) data and it turns out to be:
\begin{align*}
& \summnu < 0.59 ~\rm{ eV} \: \rm{ (95\% c.l.)} && \mbox{case (B)} \\
& \mbox{WMAP+SPT+ACT+BAO+HST} && ~ \nonumber
\end{align*}
with a best fit value of $\summnu=0.23$ eV, as we can see in Figure \ref{fig:cosmo2}.

Excluding ACT data and using only WMAP9+SPT, in case of massive neutrinos (case(B)), we obtain:
\begin{align*}
& N'_{\rm eff}  =  3.44 \pm 0.56 \: (1\sigma) && \mbox{case (B)} \\
& \mbox{WMAP+SPT} && ~ \nonumber
\end{align*}
with a total mass
\begin{align*}
& \summnu < 1.85 ~\rm{ eV} \: \rm{ (95\% c.l.)} && \mbox{case (B)} \\
& \mbox{WMAP+SPT} && ~ \nonumber
\end{align*}
In Figure \ref{fig:cosmo2} the dot--dashed (green) line shows the one dimensional marginalized posteriors for $\summnu$ and $N'_{\rm eff}$ in this case with only WMAP9+SPT. The best fit value for the sum of the masses is shifted beyond 1 eV and the range of possible mass values is much less constrained than in the above cases when ACT data are included.

\section{Joint Analysis}
\label{sec:jointanalysis}

Since the new SPT \cite{spt} and ACT \cite{act} data seem to point towards opposite directions concerning the effective number of relativistic degrees of freedom, here we adopt the conservative approach to do not combine
the two datasets in the joint analysis with SBL experiments.

Moreover, we decided to consider only the SPT dataset for the following reasons.
First of all, the ACT dataset, when only the temperature
angular power spectrum data is considered, provides an indication for gravitational lensing larger by
$\sim 70 \%$ than what expected in the standard $\Lambda$CDM model (with massless neutrinos) at more than $95 \%$ c.l.
(see \cite{act} but also \cite{divalentino}).
Since massive neutrinos decrease the gravitational lensing signal, the lensing anomaly biases the ACT results
on neutrino masses towards more stringent constraints (see Figure 3, right panel of \cite{divalentino}).
The origin of this anomalous signal is unclear and could possibly be due to a systematic in the data.
The SPT dataset, on the other hand, exhibits no anomaly in the lensing signal.

Secondly, ACT is composed by two maps in two different regions of the sky,
defined respectively as the ACT-E and the ACT-S datasets.
The ACT-S map overlaps with the region sampled by SPT and the two should
not be combined because of their covariance. The ACT-E dataset
(that constitutes approximately $50 \%$ of the entire dataset)
provides constraints that are even in larger tension with SPT (see \cite{act}).

In what follows we report the results for the joint (cosmological + SBL) analysis, obtained with different combinations of cosmological datasets. The SBL posterior probabilities on the neutrino parameters are used as priors in the MCMC sampling of those parameters relevant to cosmology, i.e. the sterile neutrino masses. In the (3+1) scheme we
have a prior on the single extra neutrino mass $m_4$; in the (3+2) scheme we have priors for
the two extra neutrino masses $m_4$ and $m_5$; finally, in (3+2) scheme we also consider
the only relevant parameter for cosmology, which is the sum of the neutrino masses $m_4+m_5$
and for that we use a prior obtained from the SBL analysis by projecting the
SBL priors for the single masses to the sum of the two.

\subsection{(3+1) Joint Analysis}
\label{subsec:3p1jointanalysis}
\begin{table*}[t]
\begin{center}
\begin{tabular}{|l|c|c|c|c|}
\hline
\hline
 Parameters & WMAP9+SPT & WMAP9+SPT & WMAP9+SPT+SDSS & WMAP9+SPT+SDSS\\
                   &                    & {\bf +SBL}             &                              & {\bf +SBL} \\
\hline
$\Omega_b h^2$ & $0.02256\pm0.00037$ &$0.02249\pm 0.00035$ & $0.02233\pm0.00034$ &$0.02230\pm 0.00034$\\
$\Omega_{dm} h^2$ & $0.131\pm0.008$ &$0.131\pm0.007$ & $0.127\pm0.007$ &$0.128\pm0.007$\\
$\theta_s$ & $1.0412\pm0.0011$ &$1.0411\pm0.0011$ & $1.0411\pm0.0011$ &$1.0412\pm0.0010$\\
$\tau$ & $0.083\pm0.013$ &$0.083\pm0.013$ & $0.081\pm0.012$ &$0.080\pm0.012$\\
$n_s$  & $0.959\pm0.011$ &$0.962\pm0.009$ & $0.963\pm0.011$ &$0.958\pm0.009$\\
$\log(10^{10} A_s)$ & $3.222\pm0.041$ &$3.218\pm0.036$ & $3.212\pm0.036$ &$3.226\pm0.034$ \\
\hline
$N_4$ & $0.65\pm0.22$ &$0.69\pm0.21$ & $<0.96$ &$<0.83$\\
$\msa [{\rm eV}]$ & $1.72\pm0.65$ &$1.27\pm0.12$ & $<2.09$ &$1.23\pm0.13$ \\
\hline
$H_0 [\rm{km/s/Mpc}]$ & $68.6\pm2.1$ &$68.9\pm1.7$ & $69.3\pm1.9$ &$68.1\pm1.4$\\
$\sigma_8$ & $0.668\pm0.061$ &$0.692\pm0.036$ & $0.766\pm0.036$ &$0.744\pm0.034$\\
$\Omega_{\rm m}$  & $0.327\pm0.035$ &$0.325\pm0.030$ & $0.311\pm0.025$ &$0.324\pm0.027$ \\
\hline
$\chi^2_{\rm min}$ & $8274.1$ &$8274.3$ & $8326.4$ &$8327.5$\\
\hline
\hline
\end{tabular}
\vspace{0.3cm}
\caption{(3+1) analysis -- Values of the cosmological parameters and their $68 \%$ confidence level intervals
in the case of one additional massive sterile neutrino, with mass $m_4$ and with multiplicity $N_4$.
The (3+1) SBL $\chi^2$ is applied where specified.
Upper bounds are quoted at $95 \%$ C.L.}
\label{tab:sbl1}
\end{center}
\end{table*}

\begin{figure}[t]
\includegraphics[scale=0.45]{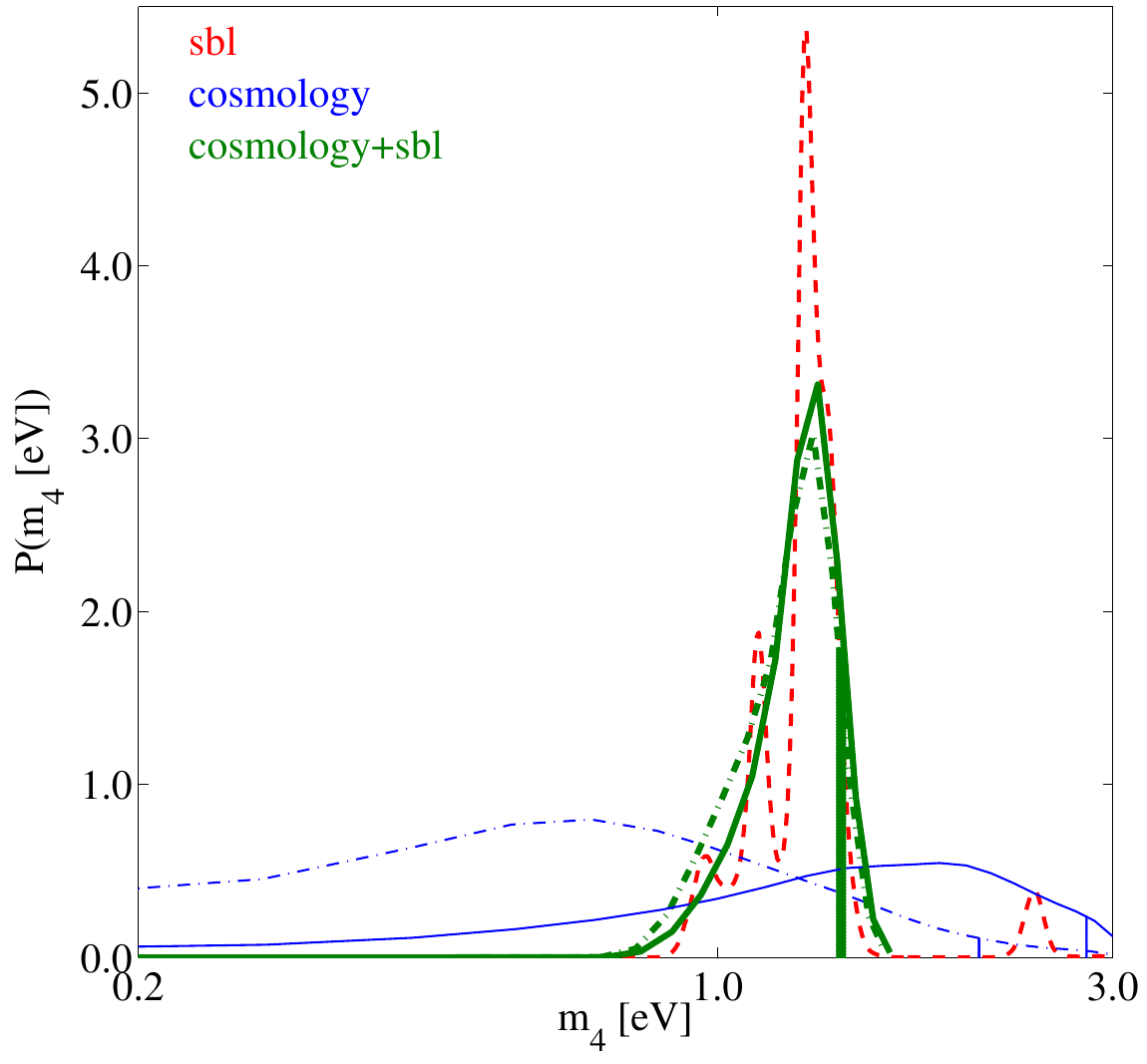}
\caption{(3+1) analysis -- One dimensional marginalized posterior for $\msa$.
The thick (green) and thin (blue) lines refer to the case of Table \ref{tab:sbl1} with and without the SBL prior, respectively.
Solid lines stand for the analysis on CMB--only (WMAP9+SPT) data;
dot--dashed lines refer to the inclusion of information from the matter power spectrum.
The (red) dashed line shows the 3+1 SBL posterior. 95\% C.L. upper bounds on the mass
for the different cases are reported as vertical lines.}
\label{fig:sbl1}
\end{figure}

\begin{figure*}[t]
\includegraphics[scale=0.45]{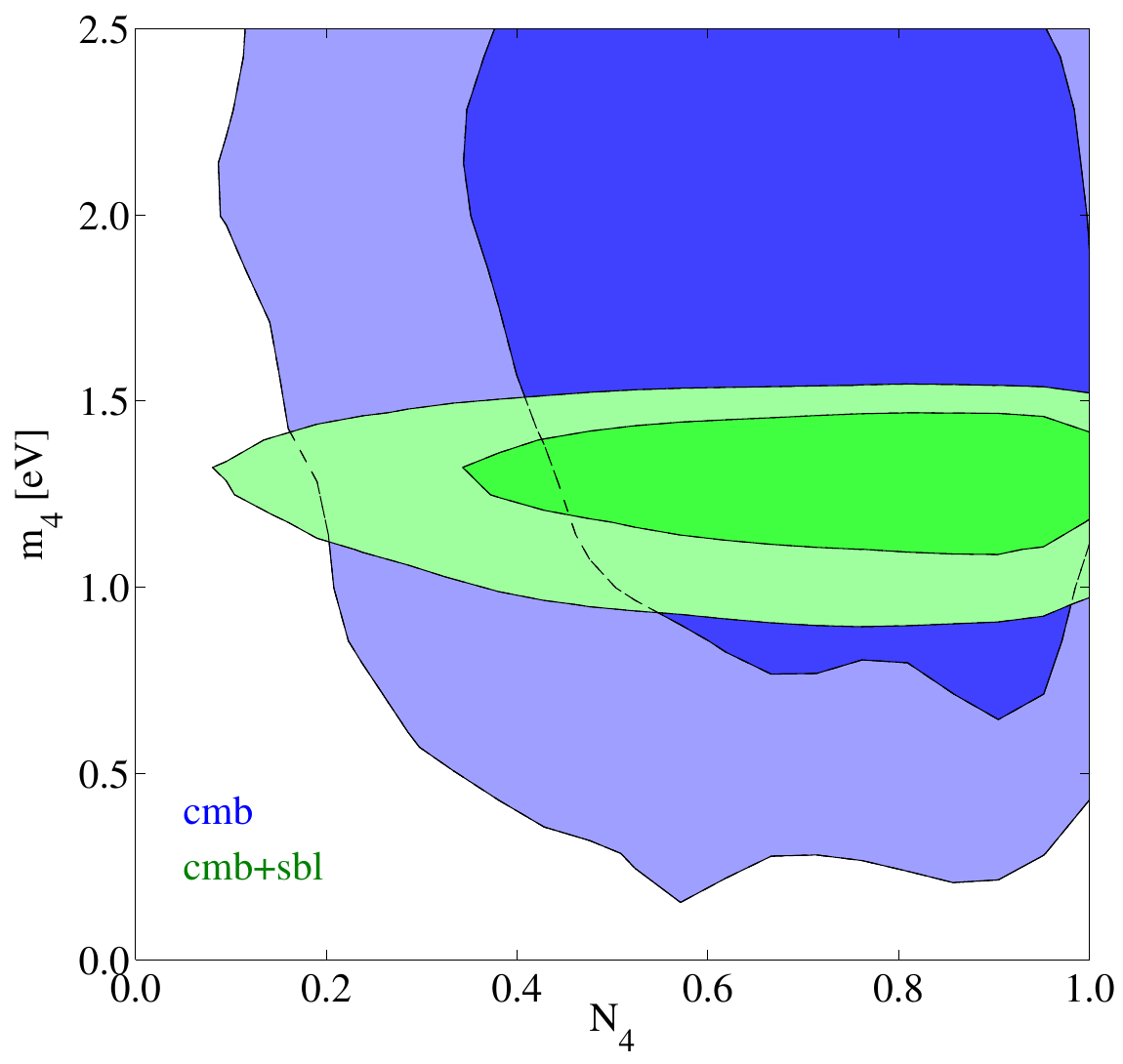}
\includegraphics[scale=0.45]{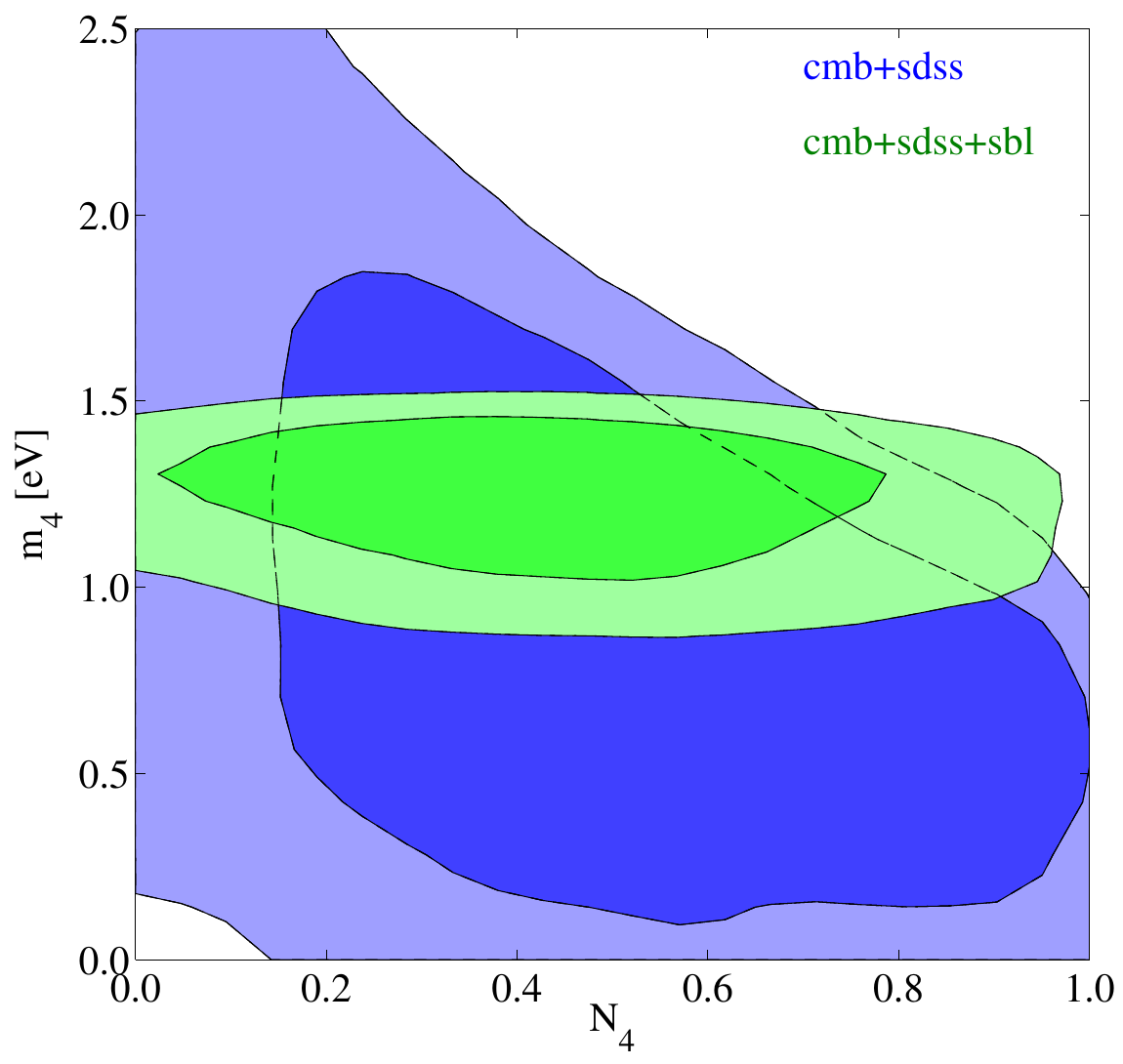}
\caption{(3+1) analysis -- Two dimensional marginalized 68\% and 95\% confidence level regions in the plane $N_4$ -- $\msa$
for the different combinations of datasets reported in Tables \ref{tab:sbl1}.}
\label{fig:sbl1bis}
\end{figure*}

The results of the joint (cosmological + SBL) analysis for the (3+1) scheme are reported in Table \ref{tab:sbl1} and in Figure \ref{fig:sbl1} and \ref{fig:sbl1bis}. The analysis refers to the case of Eq. (\ref{eq:(3+1)}). In this analysis, in addition to
the CMB datasets (WMAP9 \cite{wmap9} and SPT \cite{spt}) we also consider information on the matter
power spectrum coming from large scale observables: we include data from the Sloan Digital
Sky Survey Data Release 7 (SDSS) \cite{red}.

When only cosmological data is used, bounds on $N_4$ and $m_4$ are broad: the allowance for an
extra neutrino which cosmologically acts only as a fraction of a fully thermalized one is compatible
with cosmological data for a neutrino mass as large as 2.09 eV. The bound on the multiplicity
parameter $N_4$ is as large as 0.96. These results stand when also SDSS data are included: CMB--only data, instead, have a preference for lower values of $N_4$ and $m_4$, with best--fit values different
from zero at about the $3\sigma$ level. The central value for $N_4$ is 0.65 and deviates from zero at
$2.9\sigma$; the preferred value for $m_4$ is 1.72 eV, and differs from zero at the $2.7\sigma$ level.

When the SBL posterior probability on $m_4$ in used as a prior in the cosmological analysis, the
situations moves toward a clear preference for a neutrino mass above 1 eV and slightly lower values
for the multiplicity parameter. The result on the mass is expected, since SBL data have a clear preference
for a non-zero neutrino mass, and this reflects on the posterior probabilities of the joint analysis. The combination of CMB--only data with SBL results does not change appreciably the multiplicity
parameter (preferred value of 0.69, different from zero at the $3\sigma$ level) while it slightly
reduces the best--fit value for the mass: 1.27 eV, with an error of 0.12. The inclusion of SDSS data
enlarges the allowed interval for the multiplicity parameter, while the preferred value for the masses is further decreased to 1.23 eV, with an error or
0.13, incompatible with zero at high significance.

Figure \ref{fig:sbl1} shows the one dimensional marginalized posteriors of the joint analyses
for $\msa$ in all the cases reported in Table \ref{tab:sbl1}, together with the SBL posterior of the
(3+1) analysis.
Concerning the SBL posterior we can appreciate that the zero mass region is highly excluded and the maximum probability is characterized by three peaks between 1 eV and 2 eV, plus a small peak close to 3 eV. In the combined analysis (which includes the SBL posterior as a prior), the joint posterior closely follows the SBL posterior, with basically the same best--fit located at $\msa=1.27$ eV. The only relevant difference of the joint posterior with respect to the SBL posterior is the suppression in the joint posterior of the SBL high--mass peak close to 3 eV, that is in fact highly disfavoured by cosmological data, especially when the matter power spectrum is included in the analysis.
We notice also that the inclusion of SBL information alleviates the tension between CMB and matter power spectrum: in fact, CMB--only data exhibit a best fit value close to 2 eV and a detection of the single mass eigenstate $\msa=(1.72\pm0.65)$ eV, while the inclusion of SDSS data shifts the results towards lower values of $\msa$ and is consistent with a zero value of the mass eigenstate.

As discussed above, the value of $N_4$ is almost unconstrained when SDSS data are included,
while in the case of  CMB--only there is evidence for an extra massive sterile neutrino. This
is manifest also in Figure \ref{fig:sbl1bis}, where the the two dimensional marginalized 68\% and 95\% C.L. regions in the plane ($N_4, \msa$) are plotted for the different combinations of datasets reported in Tables \ref{tab:sbl1}. We can clearly see the effect of the SBL data: the constraints on the mass are strongly tightened but there is almost no effect on $N_4$.
Concerning the degeneracy between the number of sterile states and their mass, a negative correlation emerges when the matter power spectrum is taken into account and the result is a divergence of $\msa$ as $\nnusa$ is approaching zero.

\subsection{Bi--dimensional (3+2) Joint Analysis}
\label{subsec:3p2bijointanalysis}

\begin{table*}[t]
\begin{center}
\begin{tabular}{|l|c|c|c|c|}
\hline
\hline
 Parameters & WMAP9+SPT & WMAP9+SPT & WMAP9+SPT+SDSS & WMAP9+SPT+SDSS\\
                   &                     & {\bf +SBL}            &                              &  {\bf +SBL}                     \\
\hline
$\Omega_b h^2$ & $0.02270\pm0.00038$ &$0.02267\pm 0.00037$ & $0.02242\pm0.00036$ &$0.02235\pm 0.00035$ \\
$\Omega_{dm} h^2$ & $0.139\pm0.010$ &$0.139\pm0.009$ & $0.133\pm0.008$ &$0.131\pm0.008$ \\
$\theta_s$ & $1.0408\pm0.0011$ &$1.0409\pm0.0011$ & $1.0406\pm0.0011$ &$1.0409\pm0.0011$ \\
$\tau$ & $0.083\pm0.013$ &$0.083\pm0.013$ & $0.082\pm0.013$ &$0.080\pm0.012$ \\
$n_s$  & $0.961\pm0.012$ &$0.960\pm0.010$ & $0.967\pm0.012$ &$0.960\pm0.010$ \\
$\log(10^{10} A_s)$ & $3.228\pm0.041$ &$3.232\pm0.038$ & $3.213\pm0.037$ &$3.225\pm0.034$ \\
\hline
$N_4$ & $<0.92$ &$<0.92$ & $<0.92$ &$<0.85$ \\
$\msa [{\rm eV}]$ & $<2.11$ &$1.20\pm0.30$ & $<1.34$ &$0.95\pm0.30$\\
$N_5$  & $<0.93$ &$<0.91$ & $<0.78$ &$<0.62$ \\
$\msb [{\rm eV}]$ & $2.07\pm0.58$ &$1.96\pm0.48$ & $1.45\pm0.67$ &$1.59\pm0.49$ \\
\hline
$H_0 [\rm{km/s/Mpc}]$ & $68.5\pm2.3$ &$67.9\pm1.9$ & $69.6\pm2.1$ &$68.2\pm1.5$ \\
$\sigma_8$ & $0.642\pm0.060$ &$0.648\pm0.037$ & $0.752\pm0.036$ &$0.734\pm0.032$ \\
$\Omega_{\rm m}$  & $0.346\pm0.038$ &$0.352\pm0.037$ & $0.321\pm0.027$ &$0.331\pm0.026$ \\
\hline
$\chi^2_{\rm min}$ & $8274.1$ &$8274.8$ & $8326.4$ &$8327.8$ \\
\hline
\hline
\end{tabular}
\vspace{0.3cm}
\caption{(3+2) analysis -- Values of the cosmological parameters and their $68 \%$ confidence level intervals
in the case of two additional massive sterile neutrinos, with masses $\msa$ and $\msb$ multiplicities
$N_4$ and $N_5$.
The bi--dimensional (3+2) SBL $\chi^2$ is applied where specified.
Upper bounds are quoted at $95 \%$ C.L.}
\label{tab:sblbid}
\end{center}
\end{table*}

\begin{figure*}[t]
\includegraphics[scale=0.45]{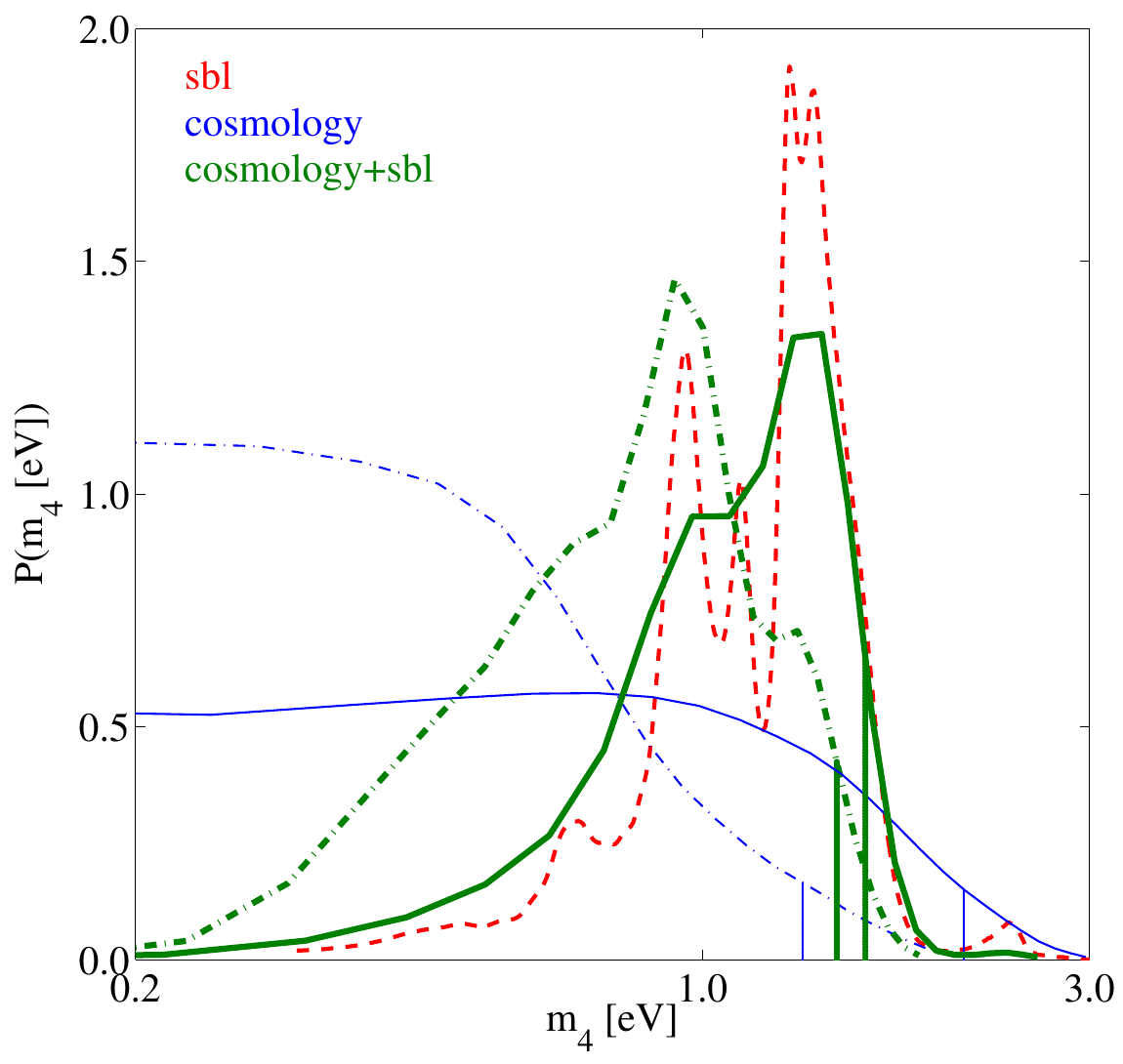}
\includegraphics[scale=0.45]{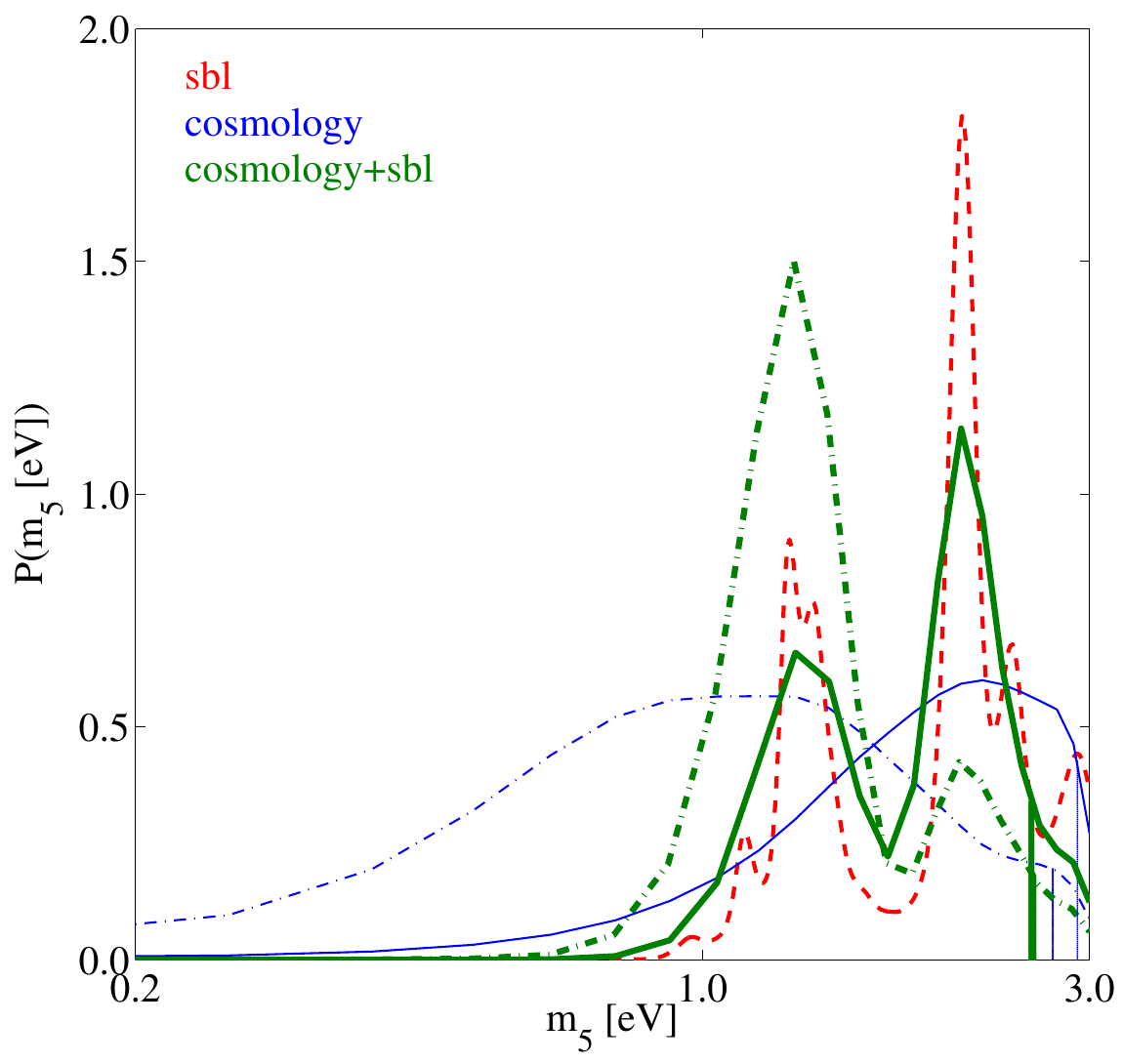}
\caption{(3+2) analysis -- One dimensional marginalized posterior probabilities for $\msa$ (left panel) and $\msb$ (right panel).
The thick (green) and thin (blue) lines refer to the case of Table \ref{tab:sblbid} with and without SBL prior, respectively.
Solid lines stand for the analysis on CMB--only (WMAP9+SPT) data;
dot--dashed lines refer to the inclusion of information coming from
the matter power spectrum. The (red) dashed line shows the bi--dimensional 3+2 SBL posterior marginalized over $\msa$ or $\msb$. 95\% C.L. upper bounds on the masses are reported as vertical lines.}
\label{fig:sblbid}
\end{figure*}

\begin{figure*}[t]
\includegraphics[scale=0.45]{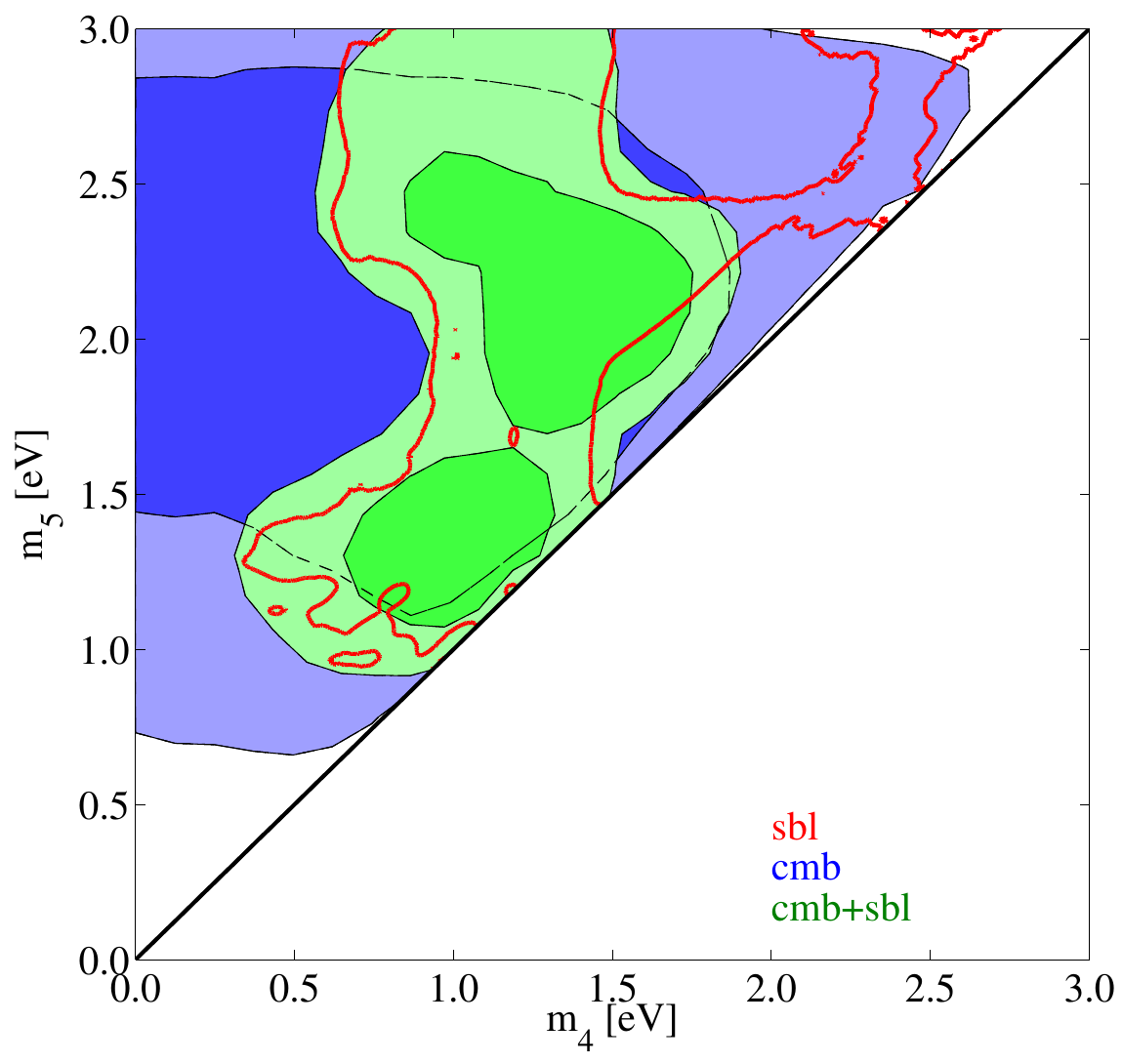}
\includegraphics[scale=0.45]{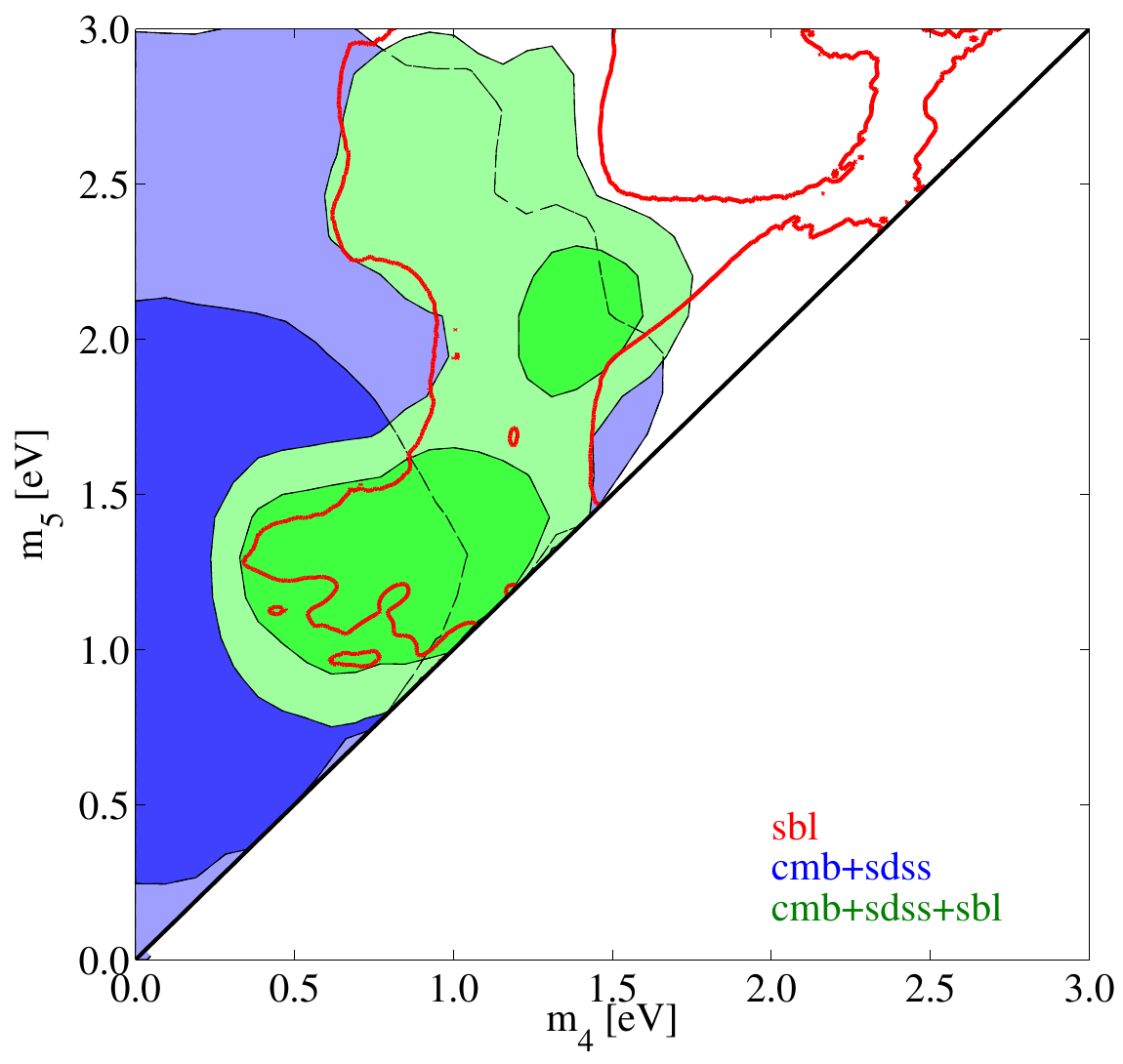}
\caption{(3+2) analysis --Two dimensional marginalized 68\% and 95\% confidence level regions in the plane ($\msa-\msb$)
for the different combinations of datasets reported in Table \ref{tab:sblbid}.
The two dimensional marginalized 95\% SBL posterior probability of the 3+2 model is also overplotted (red solid line).}
\label{fig:sblbid2}
\end{figure*}

The results of the joint analysis for the (3+2) scheme are reported in Table \ref{tab:sblbid} and in Figure \ref{fig:sblbid} and \ref{fig:sblbid2}. The analysis refers to the case of Eq. (\ref{eq:(3+2)}).

Concerning the masses, the value of the heaviest mass eigenstate  ($m_5$) is always significantly deviating from zero: when CMB--only data are considered the effect is at the $3.5\sigma$  when
the SBL are not included, and grows to a $4\sigma$ effect when SBL priors are considered. In
both cases, the neutrino mass is around 2 eV. When large--scale--structure data are included, the
preferred value for the mass decreases around 1.5 eV, with a significance of 2.1(3.2)$\sigma$
without (with) the inclusion of SBL data. SBL results, which have a clear preference for massive
neutrinos close to 1.6 eV, reinforce the cosmological results and therefore induce a clear
increase in the confidence for the mass determination.
This fact occurs also for the lighter mass eigenstate ($m_5$). While cosmological data do not require the
lightest neutrino to be massive (bounds are at 2.11 eV for CMB--only data and 1.34 eV when SDSS
is included), SBL priors induce a clear preference also for non--zero lightest mass eigenstate
around 1 eV (an effect close to the $4\sigma$ level).

The multiplicity parameter of each mass eigenstate is basically unconstrained for all the different combinations of datasets. Full and standard thermalization ($N_i=1, i=3,4$) is not allowed, but
fractional occupations as large as 0.92 are possible. A notable exception occurs for the heaviest
eigenstate ($m_5$) when SDSS data are included: in this case, the multiplicity parameter needs
to be smaller, not exceeding 0.78 for the full combination of cosmological data and further
decreasing to 0.62 when the SBL information is included.
From the last column of  Table \ref{tab:sblbid} we can conclude that cosmological and
SBL data are consistent with the existence of 2 extra sterile neutrinos, provided that they are not fully thermalised.

The same results can be appreciated with more level of detail by directly looking at the behaviour of posterior probabilities. Figure \ref{fig:sblbid} shows the one dimensional marginalized posterior for $\msa$ (left panel) and $\msb$ (right panel) in all the cases reported in Table \ref{tab:sblbid},
together with the SBL posteriors of the (3+2) scheme, marginalized over the parameters not
shown in the figure. The SBL posteriors are characterized by the clear exclusion of  zero values for both $\msa$ and $\msb$.
The cosmological posteriors of $\msa$ prefer a zero mass value, even more so when the matter power spectrum information is included in the analysis. When the SBL $\chi^2$ information is applied, the
null values for the neutrino masses become clearly disfavoured. The matter power spectrum effect turns out in a shift of the preferred mass range towards lower values.

Figure \ref{fig:sblbid2} shows the two dimensional marginalized 68\% and 95\% confidence regions in the plane $\msa$ -- $\msb$ for the different combinations of datasets reported in Table \ref{tab:sblbid},
plus the two dimensional marginalized 95\% SBL posterior of the 3+2 model.
Even when the matter power spectrum is included in the cosmological analysis, the blue contours referring to cosmological--data only show a good agreement with the red $2\sigma$ confidence region obtained by the SBL analysis.

\subsection{Reduced (3+2) Joint Analysis}
\label{subsec:3p2onejointanalysis}
\begin{table*}[t]
\begin{center}
\begin{tabular}{|l|c|c|c|c|}
\hline
\hline
 Parameters & WMAP9+SPT & WMAP9+SPT & WMAP9+SPT+SDSS & WMAP9+SPT+SDSS\\
                   &                    & {\bf +SBL}             &                              &  {\bf +SBL}                       \\
\hline
$\Omega_b h^2$ & $0.02270\pm0.00038$ &$0.02270\pm 0.00039$ & $0.02242\pm0.00036$ &$0.02241\pm 0.00035$\\
$\Omega_{dm} h^2$ & $0.139\pm0.010$ &$0.139\pm0.009$ & $0.133\pm0.008$ &$0.132\pm0.008$\\
$\theta_s$ & $1.0408\pm0.0011$ &$1.0407\pm0.0011$ & $1.0406\pm0.0011$ &$1.0407\pm0.0011$\\
$\tau$ & $0.083\pm0.013$ &$0.084\pm0.013$ & $0.082\pm0.013$ &$0.081\pm0.012$\\
$n_s$  & $0.961\pm0.012$ &$0.963\pm0.011$ & $0.966\pm0.012$ &$0.964\pm0.010$\\
$\log(10^{10} A_s)$ & $3.228\pm0.041$ &$3.225\pm0.040$ & $3.213\pm0.037$ &$3.217\pm0.035$ \\
\hline
$N_4+N_5$  & $0.97\pm0.31$ &$1.00\pm0.31$ & $0.75\pm0.32$ &$0.69\pm0.29$\\
$(\msa+\msb) [{\rm eV}]$ & $3.04\pm0.97$ &$2.80\pm0.71$ & $1.98\pm0.89$ &$2.33\pm0.61$ \\
\hline
$H_0 [\rm{km/s/Mpc}]$ & $68.5\pm2.3$ &$68.7\pm2.2$ & $69.6\pm2.1$ &$69.1\pm1.8$\\
$\sigma_8$ & $0.642\pm0.060$ &$0.647\pm0.054$ & $0.752\pm0.036$ &$0.744\pm0.033$\\
$\Omega_{\rm m}$  & $0.346\pm0.038$ &$0.344\pm0.037$ & $0.321\pm0.027$ &$0.325\pm0.026$ \\
\hline
$\chi^2_{\rm min}$ & $8274.1$ &$8275.1$ & $8326.4$ &$8327.6$\\
\hline
\hline
\end{tabular}
\vspace{0.3cm}
\caption{Reduced (3+2) analysis -- Values of the cosmological parameters and their $68 \%$ confidence level intervals, for the data analyzed in terms of of the sum of the sterile neutrino mass
eigenstates  $(m_4+m_5)$ and of their total effective multiplivity $N_4+N_5$.
The one--dimensional (3+2) SBL $\chi^2$ is applied where specified.
Upper bounds are quoted at $95 \%$ C.L.}
\label{tab:sbl2}
\end{center}
\end{table*}

\begin{figure}[t]
\includegraphics[scale=0.45]{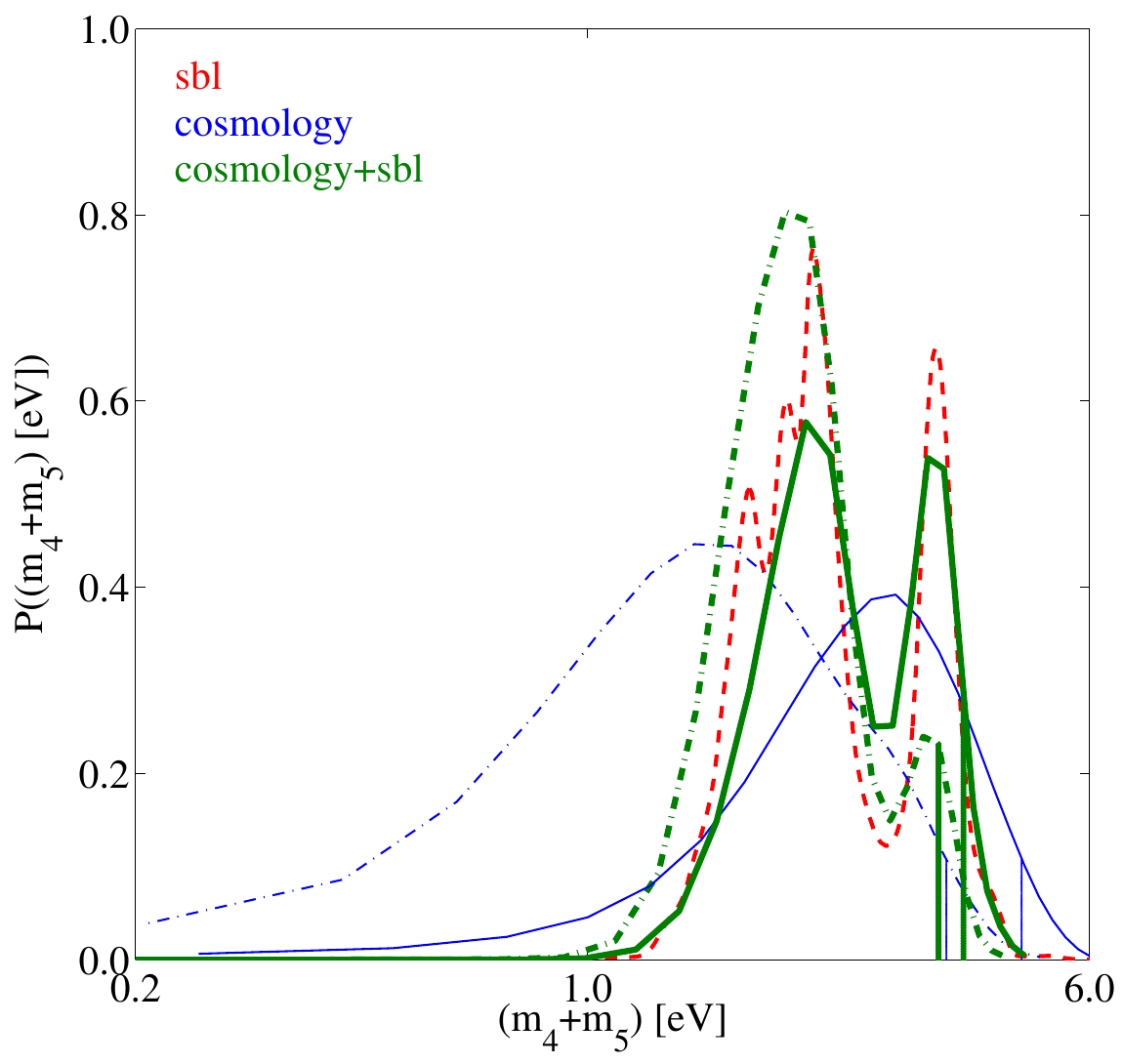}
\caption{Reduced (3+2) analysis -- One dimensional marginalized posterior probabilities for $(\msa+\msb)$.
The thick (green) and thin (blue) lines refer to the case of Table \ref{tab:sbl2} with and without SBL prior, respectively.
Solid lines stand for the analysis of CMB--only  (WMAP9+SPT) data;
 dot--dashed refer to the inclusion of information from
 the matter power spectrum.
 The (red) dashed line shows the one--dimensional 3+2 SBL posterior on $(\msa+\msb)$.
95\% C.L. upper bounds are also reported (vertical lines).
}
\label{fig:sbl2}
\end{figure}

\begin{figure*}[t]
\includegraphics[scale=0.45]{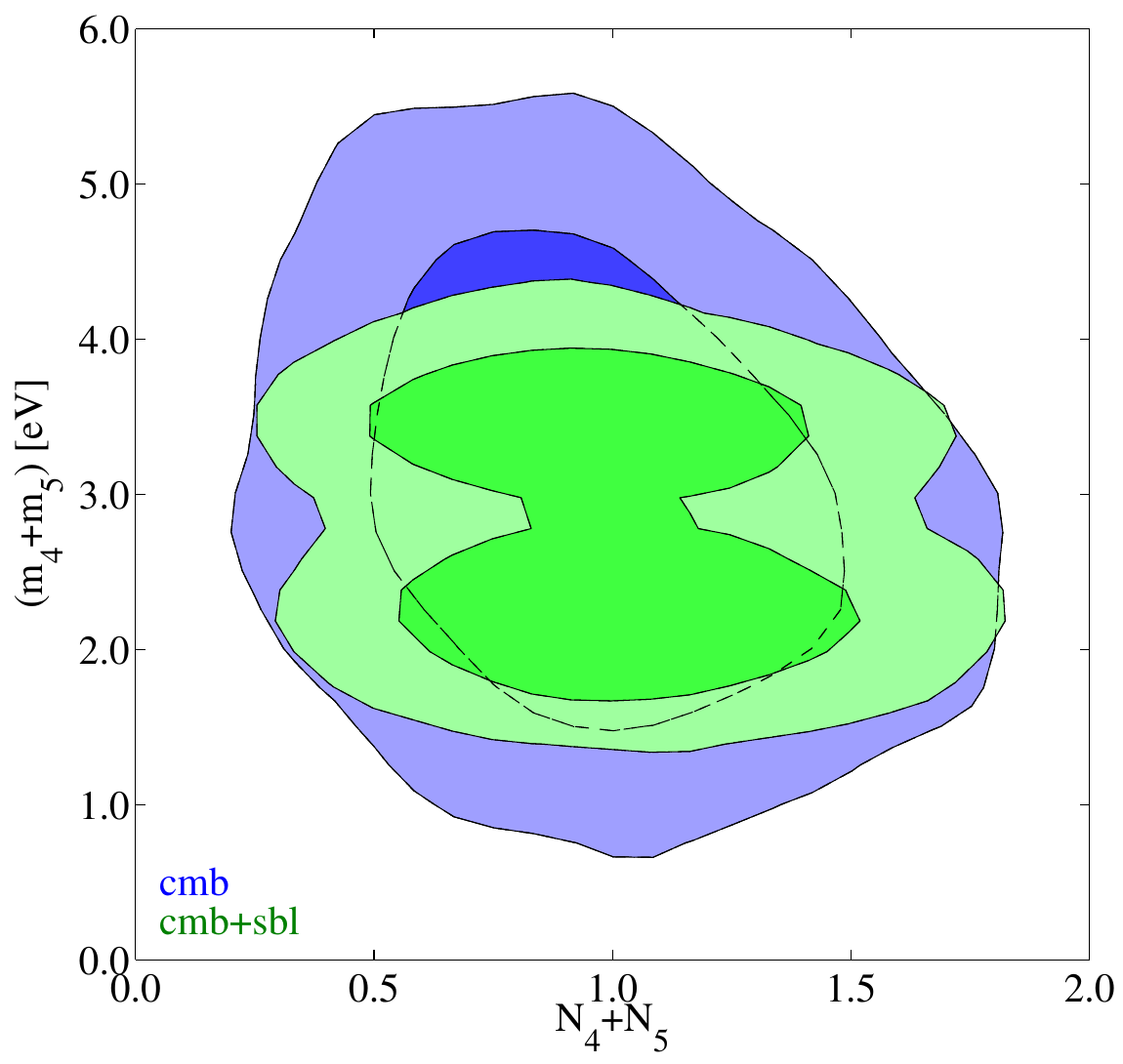}
\includegraphics[scale=0.45]{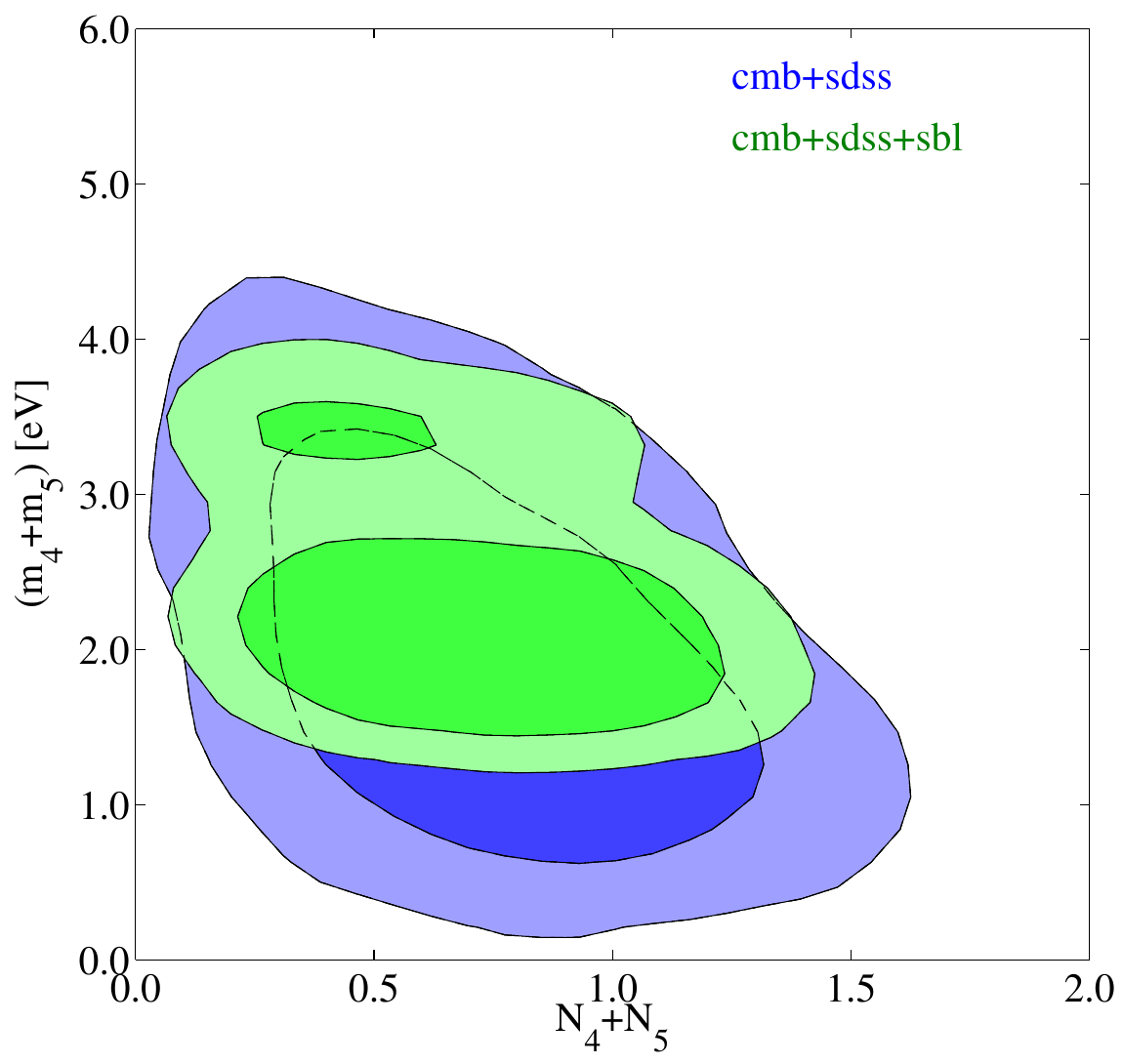}
\caption{Reduced (3+2) analysis -- Two dimensional marginalized 68\% and 95\% confidence level regions in the plane $(N_4+N_5)$ vs. $(\msa+\msb)$
for the different combinations of datasets reported in Table \ref{tab:sbl2}.}
\label{fig:sbl2bis}
\end{figure*}

Finally, in Table \ref{tab:sbl2} and in Figure \ref{fig:sbl2} and \ref{fig:sbl2bis} we show the results of
the join (cosmological + SBL) analysis in the (3+2) scheme, where we have post--processed the chains of the MCMC to get constraints only on the sum of the mass eigenstates and on the total number of extra species, by applying a SBL prior on $(\msa+\msb)$ obtained by realizing a one--dimensional SBL posterior on the sum of the two masses.

Interestingly in this case the sum of the neutrino masses always shows a clear preference for a non zero value, even when only cosmological data are considered: the confidence level for CMB--only
data is $3.1\sigma$, and degrades to $2.2\sigma$ when SDSS data are included. The inclusion
of the SBL information sizably strengthen the result, driving the confidence of non--zero masses
close to a $4\sigma$ level.

Figure \ref{fig:sbl2} shows these behaviours by reporting  the one dimensional marginalized posterior for $\msa+\msb$ in all the cases reported in Table \ref{tab:sbl2}, together with the posteriors obtained by analyzing the SBL data within the framework of the (3+2) model and marginalizing over $\msa+\msb$.
The situation is analogous to the one we found in Section \ref{subsec:3p2bijointanalysis}. The cosmological posteriors alone show a preference for a non--zero value for the sum of the masses. As usual, when SDSS data are considered, the posterior is shifted towards lower values for the sum of the masses, but the preference remains for masses around 1 eV. Concerning the SBL posterior, the maximum probability on the sum of the masses is around 2 eV as expected, with a tail at higher masses. This higher mass tail is suppressed in the joint posterior, when the cosmological analysis considers also the matter power spectrum information. Instead, a higher mass tail can be seen in the joint posterior when the combined analysis takes into account CMB--only data as cosmological datasets.

Figure \ref{fig:sbl2bis} shows the two dimensional marginalized 68\% and 95\% confidence level regions in the plane $(N_4+N_5)$ vs. $(\msa+\msb)$, for the different combinations of datasets reported in Tables \ref{tab:sbl2}. Once again, we can clearly see the effect of the SBL data: the constraints on the sum of the masses are tightened but there is almost no effect on the number of extra massive sterile neutrino species. Furthermore, in Figure \ref{fig:sbl2bis} we recognize the same degeneracy between the sum of the mass eigenstates and their multiplicity that was seen in Figure \ref{fig:sbl1bis}. As we have already discussed, this degeneracy appears when matter power spectrum information is taken into account. Interestingly, here this degeneracy remains also when the SBL posterior is applied.

\section{Conclusions}
\label{sec:conclusions}

The case for Extra Dark Radiation in cosmology has recently become very complex and somewhat controversial:
the new WMAP9 and SPT data are still pointing toward an extra component while the new ACT results became compatible with the standard cosmological value $\neff=3.046$.
On the other hand short--baseline neutrino oscillation experiments keep confirming the LSND anomaly and the presence of one or two sterile neutrinos.

In this framework, our analysis provides an update of the cosmological results in the sterile neutrino context, taking into account all the present CMB data (WMAP9, SPT, ACT) and investigating all the possible parameterizations (massless; massive; 3 active massless plus a varying number of massive sterile states). 

We find that even in the context of such very general models the ACT and SPT data are not compatible, and that the results are strongly affected by this discrepancy. This in turn affects the inferred neutrino mixing and thermalization parameters.
The discrepancy is to some extent alleviated when BAO data plus a prior on $H_0$ from the HST analysis are included: in this case the effective number
of (massive) neutrinos is $3.60\pm0.35$ with a total effective mass 0.59 eV.

Given this discrepancy we have not used the ACT dataset in the rest of our analysis (because it also leads to spuriously stringent bounds on the neutrino mass), and we considered only WMAP9 and SPT in combination with SDSS--DR7.
In order to analyze the cosmological evidences in the context of neutrino mass models and short--baseline neutrino oscillation data, we have used the results from the SBL as a prior in the analyses of
cosmological data. We specifically considered models with one or two extra sterile neutrinos (denoted
by (3+1) and (3+2) schemes): we first performed a full analysis of the SBL data and then we used the
SBL posterior probabilities on the sterile neutrino masses as priors in the MCMC analysis of
the cosmological datasets.

We found that the inclusion of the SBL priors induce tight bounds on the sterile neutrino mass eigenstates, and
mildly constrain the fractional contribution of the extra neutrinos to the total energy budget. In the (3+1) scheme we obtain $\msa=(1.27\pm0.12)$ eV when CMB--only data are considered as cosmological datasets, and $\msa=(1.23\pm0.13)$ eV when SDSS--DR7 data are also included. We also notice that
there is an evidence for a non zero value of the single mass eigenstate (although with a larger uncertainty) even without SBL priors when only CMB data are considered: $\msa=(1.72\pm0.65)$ eV.
In the (3+1) scheme, the inclusion of SBL information does not significantly constrain the multiplicity
parameter, which could be as large as 0.96. Instead, if SDSS data are not included in the analysis, the multiplicity parameter can deviate from zero at about $3\sigma$.

In the (3+2) context, the inclusion of SBL information induces relatively tight intervals for the two
mass eigenstates: $\msa=(1.20\pm0.30)$ eV and $\msb=(1.96\pm0.48)$ eV for CMB--only data;
$\msa=(0.95\pm0.30)$ eV and $\msb=(1.59\pm0.49)$ eV when SDSS--DR7 is also included. In
this scheme, the multiplicity parameters are again only slightly bounded from above, except for
the heaviest mass eigenstate: when SDSS data are included, $N_5$ cannot exceed 0.62, meaning
that the sterile neutrino need to be only partially contributing to the energy density of the Universe.

In conclusion, we found that the SBL data exhibit a good agreement with the new updated cosmological bounds on neutrino masses: this occurs both when considering CMB--only data
(WMAP+SPT) and when adding information from SDSS. Either one or two extra sterile neutrinos,
with a mass at (or even above) the eV scale (as dictated by the SBL results) and not fully contributing to the energy density (as can occur, e.g., in the case of partial thermalization) are therefore fully compatible with current cosmological measurements.

\section{Planck results}
\label{sec:planck}

A few weeks after the completion of this paper, the Planck Collaboration has released its results \cite{Ade:2013lta}. Concerning the effective number of relativistic degrees of freedom, Planck data combined with WMAP-9 polarization data and high-l data (both ACT and SPT) found $\neff$ consistent with the standard value within $1\sigma$. This result is deeply affected by the Planck measure of the Hubble constant: if the HST prior on $H_0$ is considered, Planck results recover a $2\sigma$ evidence of dark radiation, with an evidence slightely weaker if BAO data are included. The impact of Planck data on dark radiation results strongly depend on the additional dataset considered in the analyses.
Furthermore even considering the most stringent results, a mild evidence of an extra component of the radiation content of the Universe is still present in the latest data.

\begin{figure*}[t]
\includegraphics[scale=0.45]{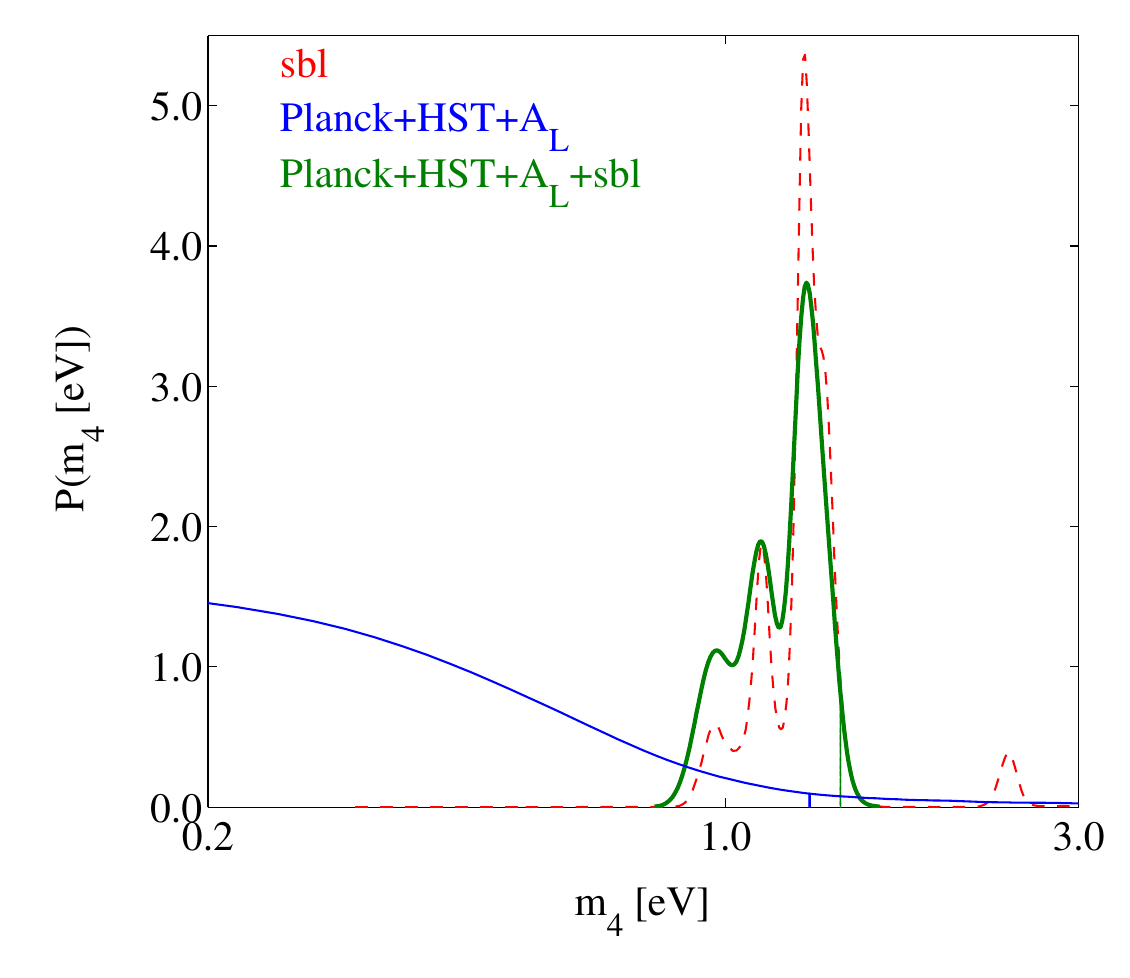}
\caption{(3+1) analysis -- One dimensional marginalized posterior for $\msa$.
The thick (green) and thin (blue) lines refer to the case of Table \ref{tab:sbl1pl} with and without the SBL prior, respectively.
The (red) dashed line shows the 3+1 SBL posterior. 95\% C.L. upper bounds on the mass
for the different cases are reported as vertical lines.}
\label{fig:sbl1pl}
\end{figure*}

\begin{figure*}[t]
\includegraphics[scale=0.45]{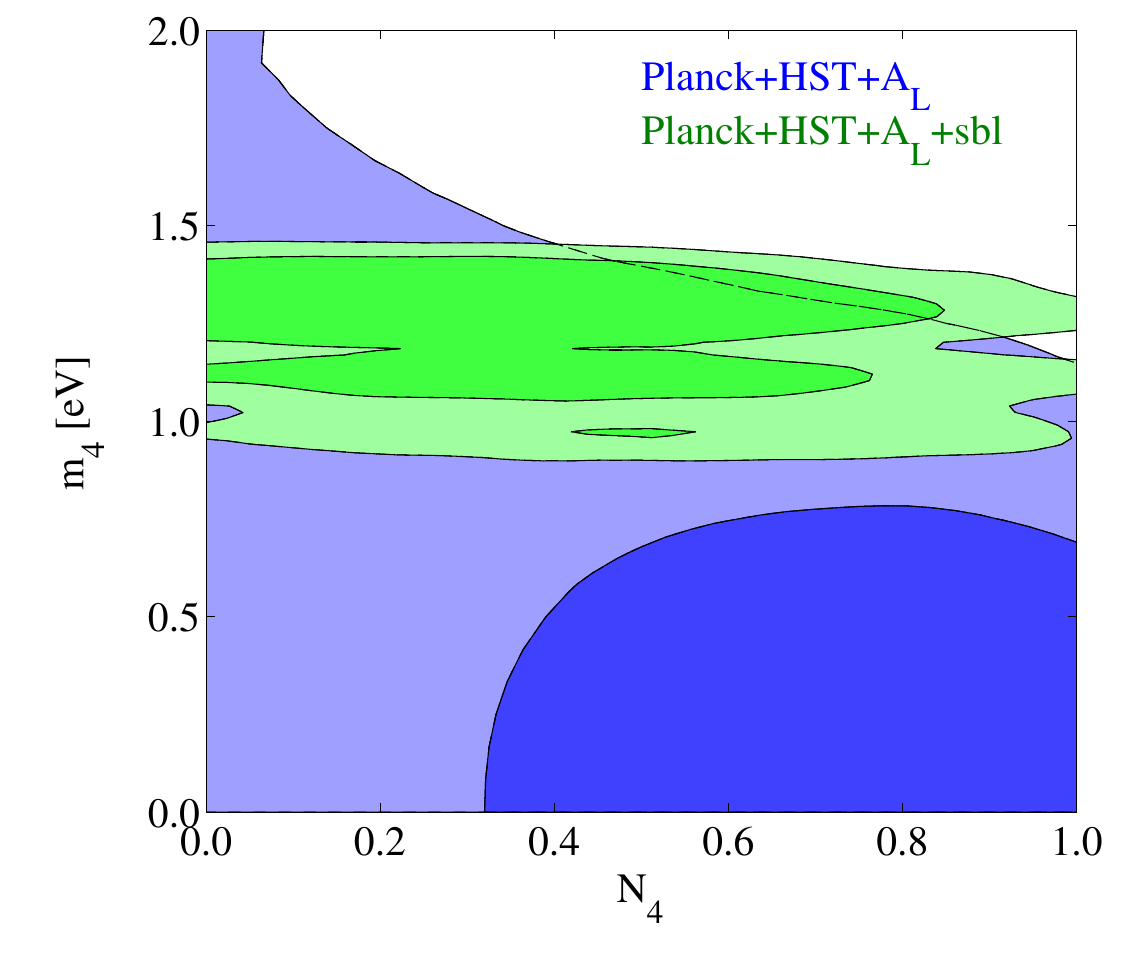}
\caption{(3+1) analysis -- Two dimensional marginalized 68\% and 95\% confidence level regions in the plane $N_4$ -- $\msa$.}
\label{fig:sbl1plbis}
\end{figure*}

In order to check the consistency of our conclusions with the latest CMB data, we have applied our 3+1 model to Planck data, including also WMAP-9 polarization and HST prior on $H_0$. In this case we also let the lensing amplitude $A_{\rm L}$ free to vary accordingly to the definition given in \cite{Calabrese:2008rt}. As clearly showed in \cite{Ade:2013lta}, the inclusion of
a variation in the lensing amplitude of the temperature spectrum is needed in order to avoid biased results on the
neutrino mass.  Our choice of including the lensing amplitude is indeed motivated by the higher sensitivity of Planck: now the dominant effect of massive neutrinos is related to the gravitational lensing rather than to the early-Integrated Sachs Wolfe effect.
In Table \ref{tab:sbl1pl} the new results are reported and it is clear that they still allow for an interpretation of an extra dark radiation component in terms of partially thermalized sterile neutrinos: $N_4$ turns out to be almost uncostrained and the 95\% upper bound on $\msa$ is consistent with the mass region allowed by SBL posterior (thin (blue) solid line and red dotted line, respectively, in Figure \ref{fig:sbl1pl}). As a consequence the joint analysis (thick (green) solid line in Figure \ref{fig:sbl1pl})) is still acceptable and the joint posterior perfectly matches the SBL.
Figure \ref{fig:sbl1plbis} shows that there is no degeneracy among $N_4$ and $\msa$ even when only cosmological data are included and the SBL prior is taken out. The reason is that Planck accuracy makes possible resolving neutrino mass and neutrino number separately.
Finally we can appreciate the effect of the SBL prior on the cosmological data.
This effect concerns only the sterile mass eigenstate $\msa$ but it doesn't affect the constraints on the multiplicity $N_4$.
The results presented in this paper are therefore compatible with the new Planck results. We will present a more detailed
analysis of the Planck data, in combination with other datasets, in a forthcoming paper \cite{marianew}.

\begin{table*}[t]
\begin{center}
\begin{tabular}{|l|c|c|}
\hline
\hline
 Parameters & Planck+HST & Planck+HST\\
                   &                    & {\bf +SBL} \\
\hline
$\Omega_b h^2$ & $0.02287\pm0.00035$ &$0.02290\pm 0.00037$\\
$\Omega_{c} h^2$ & $0.1226\pm0.0048$ &$0.1180\pm0.0039$\\
$\theta_s$ & $1.0408\pm0.0008$ &$1.0414\pm0.0008$\\
$\tau$ & $0.092\pm0.014$ &$0.091\pm0.013$\\
$n_s$  & $0.986\pm0.010$ &$0.979\pm0.009$\\
$\log(10^{10} A_s)$ & $3.144\pm0.037$ &$3.157\pm0.034$\\
\hline
$N_4$ & $0.62\pm0.25$ &$0.42\pm0.25$\\
$\msa [{\rm eV}]$ & $<1.64$ &$1.21\pm0.14$\\
$A_{\rm L}$ & $1.29\pm0.13$ & $1.34\pm0.13$ \\
\hline
$H_0 [\rm{km/s/Mpc}]$ & $72.1\pm2.0$ &$70.0\pm1.3$ \\
$\sigma_8$ & $0.793\pm0.041$ &$0.741\pm0.042$ \\
$\Omega_{\rm m}$  & $0.286\pm0.019$ &$0.299\pm0.021$ \\
\hline
\hline
\end{tabular}
\vspace{0.3cm}
\caption{(3+1) analysis -- Values of the cosmological parameters and their $68 \%$ confidence level intervals
in the case of one additional massive sterile neutrino, with mass $m_4$ and with multiplicity $N_4$.
The (3+1) SBL $\chi^2$ is applied where specified.
Upper bounds are quoted at $95 \%$ C.L.}
\label{tab:sbl1pl}
\end{center}
\end{table*}

\begin{acknowledgements}
MA acknowledges the European ITN project Invisibles (FP7-PEOPLE-2011-ITN, PITN-GA-2011-289442-INVISIBLES).
NF and CG acknowledge acknowledge INFN research grant FA51.
NF acknowledges support of the spanish MICINN
Consolider Ingenio 2010 Programme under grant MULTIDARK CSD2009- 00064 (MICINN).
\end{acknowledgements}

\end{document}